\documentclass[bibyear]{aa}

\usepackage{graphicx}
\usepackage{txfonts}
\usepackage{natbib}
\usepackage{lipsum}
\usepackage[bbgreekl]{mathbbol}

\usepackage[colorlinks=true, linkcolor=blue, citecolor=blue, bookmarks=false]{hyperref}

\bibpunct{(}{)}{;}{a}{}{,} 

\newcommand{\bt}{\boldsymbol{\theta}}
\newcommand{\bvt}{\boldsymbol{\vartheta}}
\newcommand{\btt}{\tilde{\boldsymbol{\theta}}}

\newcommand{\bts}{\boldsymbol{\theta}^{\star}}

\newcommand{\tA}{\theta_{\text{\scalebox{.9}{A}}}}
\newcommand{\tB}{\theta_{\text{\scalebox{.9}{B}}}}
\newcommand{\tC}{\theta_{\text{\scalebox{.9}{C}}}}
\newcommand{\tI}{\theta_{\text{\scalebox{.9}{I}}}}
\newcommand{\ttt}{\theta_{\text{t}}}
\newcommand{\tr}{\theta_{\text{r}}}
\newcommand{\tE}{\theta_{\text{\scalebox{.8}{E}}}}
\newcommand{\tc}{\theta_{\text{\tiny{c}}}}

\newcommand{\ba}{\boldsymbol{\alpha}}
\newcommand{\bha}{\hat{\ba}}
\newcommand{\bta}{\tilde{\ba}}
\newcommand{\ha}{\hat{\alpha}}

\newcommand{\kp}{\kappa}
\newcommand{\hkp}{\hat{\kp}}
\newcommand{\tkp}{\tilde{\kp}}

\newcommand{\bb}{\boldsymbol{\beta}}
\newcommand{\bhb}{\hat{\bb}}
\newcommand{\hb}{\hat{\beta}}
\newcommand{\btb}{\tilde{\boldsymbol{\beta}}}

\newcommand{\bbs}{\boldsymbol{\beta}^{\star}}

\newcommand{\hpsi}{\hat{\psi}}
\newcommand{\tpsi}{\tilde{\psi}}

\newcommand{\htau}{\hat{\tau}}
\newcommand{\ttau}{\tilde{\tau}}
\newcommand{\hT}{\hat{T}}
\newcommand{\tT}{\tilde{T}}

\newcommand{\varbeta}{\beta_{\text{s}}}
\newcommand{\bn}{\boldsymbol{\nabla}}
\newcommand{\sub}[2]{#1_{\text{\scalebox{.9}{#2}}}}
\newcommand{\subexp}[3]{#1_{\text{\scalebox{.9}{#2}}}^{\text{\scalebox{.9}{#3}}}}
\newcommand{\gp}{\sub{\gamma}{p}}
\newcommand{\hA}{\hat{\mathcal{A}}}
\newcommand{\eAB}{\sub{\varepsilon}{AB}}
\newcommand{\nAB}{\sub{\eta}{AB}}
\newcommand{\pythonpackage}{\texttt{pySPT}}

\newcommand{\python}{\texttt{python}}

\begin{document} 

\title{Ambiguities in gravitational lens models: impact on time delays of the source position transformation}

\titlerunning{Source position transformation}

\author{Olivier Wertz\inst{1},
	    Bastian Orthen\inst{1}
            \and Peter Schneider\inst{1}
            }       

\institute{Argelander-Institut f\"ur Astronomie, Auf dem H\"ugel 71, D-53121 Bonn}

\date{Received 6 November 2017 /  Accepted 11 December 2017}

\abstract
{
The central ambition of the modern time delay cosmography consists in determining the Hubble constant $H_0$ with a competitive precision. 
However, the tension with $H_0$ obtained from the \emph{Planck} satellite for a spatially-flat $\Lambda$CDM cosmology suggests that systematic errors may have been underestimated.
The most critical one probably comes from the degeneracy existing between lens models that was first formalized by the well-known mass-sheet transformation (MST). 
In this paper, we assess to what extent the source position transformation (SPT), a more general invariance transformation which contains the MST as a special case, may affect the time delays predicted by a model.
To this aim we use \pythonpackage, a new open-source \python\ package fully dedicated to the SPT that we present in a companion paper.
For axisymmetric lenses, we find that the time delay ratios between a model and its SPT-modified counterpart
simply scale like the corresponding source position ratios, $\Delta \hat{t}/ \Delta t \approx \hb/\beta$, regardless of the mass profile and the isotropic SPT. 
Similar behavior (almost) holds for non-axisymmetric lenses in the double image regime and for opposite image pairs in the quadruple image regime. In the latter regime, we also confirm 
that the time delay ratios are not conserved. In addition to the MST effects, the SPT-modified time delays deviate in general no more than a few percent for particular image pairs, 
suggesting that its impact on time-delay cosmography seems not be as crucial as initially suspected.
We also reflected upon the 
relevance of the SPT validity criterion and present arguments suggesting that it should be reconsidered. Even though a new validity criterion would affect the time delays in a different way, 
we expect from numerical simulations that our conclusions will remain unchanged. 
 }

\keywords{cosmological parameters -- gravitational lensing: strong}

\maketitle
\titlerunning
\authorrunning

\section{Introduction}
\label{section:introduction}

The use of the gravitational lensing phenomenon 
as a cosmological tool offers an independent way to probe the nature of 
the universe \citep[for the early work see][]{Blandford_Narayan_Review_GL_and_Cosmo}.
To date, numerous weak and strong lensing observations have been employed to infer the fundamental cosmological parameters with an increasingly competitive precision. 	
In the strong lensing regime, \citet{Refsdal_1964b} established that multiple-image systems 
can theoretically be used to infer the Hubble parameter $H_0$. 
The method relies upon the idea that the propagation time of light rays emitted from a background source (typically an Active Galactic Nucleus, AGN) towards the observer 
differs from one lensed image to another. The corresponding difference in arrival times, known as the time delay, is inversely proportional to $H_0$. 
This idea lays the basis of the modern time-delay cosmography, which has been extensively addressed in literature; see for example the recent review \cite{Treu_Marshall_TD_Cosmography_review_2016} and references therein.

For sake of clarity, we recall few key results of the well-known theory of time delays. 
Relative to an unperturbed ray emitted by a source located at $\bb$, the extra light travel time $T(\bt)$ at an image position $\bt$ is 
formally defined by
\begin{equation}
	T(\bt) = \frac{D_{\Delta t}}{c} \left\{\frac{1}{2} \left[\bt - \bb(\bt)\right]^2 - \psi(\bt) \right\} \eqqcolon \frac{D_{\Delta t}}{c}\ \tau(\bt) \ ,
	\label{oTau}
\end{equation}
where $\psi(\bt)$ is the deflection potential produced by a dimensionless surface mass density $\kappa(\bt) = \bn^2 \psi(\bt)/2$, 
$\tau(\bt)$ is known as the Fermat potential, and $D_{\Delta t}$ is referred 
to as the time-delay distance 
\begin{equation}
	D_{\Delta t} = (1 + \sub{z}{d}) \frac{\sub{D}{d} \sub{D}{s}}{\sub{D}{ds}} \propto \sub{H^{-1}}{0} \ ,
	\label{TD_Distance}
\end{equation}
where $\sub{z}{d}$ is the redshift of the deflector and $D$ the angular diameter distances between the observer and deflector (d), observer and source (s), and deflector and source (ds).
In Eq.\,\eqref{oTau}, the first term in brackets describes the geometrical deviation of the light ray 
due to the lens whereas the second describes the time delay that a ray experiences as it crosses the deflection potential.
The relative time delay $\Delta t_{ij}$ between a pair of lensed images $\bt_i$ and $\bt_j$ is obtained by differencing the corresponding extra light travel time
\begin{equation}
	\Delta t_{ij} = T(\bt_i) - T(\bt_j) =  \frac{D_{\Delta t}}{c} \left[\tau(\bt_i) - \tau(\bt_j)\right] \eqqcolon \frac{D_{\Delta t}}{c}  \Delta \tau_{ij} \ .
	\label{oTD}
\end{equation}

From Eq.\,\eqref{oTD}, $\sub{H}{0}$ inference can be conceptually performed by constraining the time-delay distance $D_{\Delta t}$, provided that both accurate time
delay measurements and a mass model which predicts $\Delta \tau_{ij}$ can be obtained. At present, a few percent precision time delays have been measured for several 
multiple-image systems based on different light curve analysis methods 
\citep[see e.g.][]{	Vuissoz_TD_2008, 
				Paraficz_TD_2010, 
				Courbin_TD_2011, 
				Fohlmeister_TD_2013, 
				Eulaers_TD_2013, 
				Tewes_TD_2013, 
				RathnaKumar_TD_2013, 
				Bonvin_TD_2017,
				Akhunov_Wertz_TD_2017}.
In the foreseeable future, we can expect thousands of lensed quasars to be discovered by the next generation of instruments 
\citep{Jean_ILMT_GL_2001, Coe_Moustakas_2009, Oguri_Marshall_2010, Finet_Gaia_2012, Finet_XXL_2015, Liao_TDC_2_2015, Finet_Surdej_Gaia_2016}.
Among them, numerous suitable candidates for robust time delay measurements should lead the time delay cosmography to the next level. 
However, constraining the lens mass distribution turns out to be as decisive as measuring time delays with high precision. 
Given a measured time delay between two lensed images, more concentrated mass distributions lead to shorter time-delay distance estimations, hence to larger values of $\sub{H}{0}$ 
\citep{Kochanek_2002_TD}. 
The Fermat potential difference $\Delta \tau_{ij}$ is primarily sensitive 
to the strong lensing effects produced by the main lens. However, a realistic time delay cosmography should also consider the lensing effects of any external mass structures located in the vicinity 
of the main lens, as well as along the line of sight \citep[LOS; e.g.,][]{Seljak_1994, BarKana_1996}. If the LOS mass effects are sufficiently small, they can be approximated by an external shear and an external 
convergence, usually denoted as $\sub{\kappa}{ext}$, which need to be characterized\footnote{We note that the time delay distance $D_{\Delta t}^{\text{model}}$ inferred from a model that neglects the impact of $\sub{\kappa}{ext}$ is related to the true time 
delay distance by $D_{\Delta t}^{\text{model}} = (1 - \sub{\kappa}{ext})\ D_{\Delta t}$.} 
\citep[see e.g.][]{Keeton_substructures_2003, Fassnacht_B1608p656_2006, Suyu_2010_B1608+656_H0, Wong_2011, Suyu_timeDelayDistance_2013, Wong_HOLiCOW_2017}.
Otherwise, these external mass structures need to be explicitly included in the mass model, for instance by considering the full multi-plane lensing formalism \citep{Schneider_generalizedMultiplane_2014, McCully_2014, McCully_2016}.
 
As first shown in \cite{Falco_MST_1985}, the dimensionless surface mass density $\kappa(\bt)$ and the class of mass models $\kappa_{\lambda}(\bt)$ defined as
 \begin{equation}
	\kappa_{\lambda}(\bt) = \lambda\,\kappa(\bt) + (1 - \lambda) \ ,
	\label{MST}
\end{equation}
along with the corresponding unobservable source rescaling $\bb \rightarrow \lambda\,\bb$, lead to identical lensing observables, except for the time delays between pairs of lensed images 
which are transformed such that $\Delta t \rightarrow \lambda\,\Delta t $.
If not broken, this degeneracy, referred to as the mass-sheet transformation (MST), may significantly affect cosmographic inferences, including $\sub{H}{0}$ \citep[see e.g.][]{Gorenstein_Falco_Shapiro_1988, Saha_degeneracies_2000, Wucknitz_TD_2002, Koopmans_H0_MSD_B1608p656_2003, Liesenborgs_2012, SPT_SS13, Schneider_MST_multiplane_2014, Schneider_generalizedMultiplane_2014, Xu_Illustris_H0_2016}. 
It is worth mentioning that the external convergence $\sub{\kappa}{ext}$ is based on physical effects whereas the MST \eqref{MST} stems from a pure mathematical degeneracy \citep[][hereafter SS13]{SPT_SS13}.  
Different solutions have been proposed to reduce the degeneracy induced by the MST in time delay cosmography \citep[see e.g. \S 3 in][and references therein]{Treu_Marshall_TD_Cosmography_review_2016}. A commonly used method consists in assuming a specific lens model, typically a power-law, and independently estimating the lens mass with the measurement of its velocity dispersion. 
However, SS13 have shown experimentally that two different classes of galaxy models with compatible velocity dispersions were able to reproduce equally well a set of image positions, but predicted significantly different time delays. 
Furthermore, because the time delay ratios were not constant, they suggested that the transformation between these two models was not exactly an MST but a more general one.
This has naturally raised some concerns about the reliability of the $H_0$ determination from time delay cosmography.

\citet[hereafter SS14]{SPT_SS14} laid the theoretical basis for an approximate invariance transformation, the so-called source-position transformation (SPT), 
of which the MST is a special case. 
\citet[][hereafter USS17]{SPT_USS17} explored further its properties, such as defining a criterion to determine whether an SPT is valid or not 
and exploring the density profile of SPT-modified mass distributions. 
They also pointed out that the degeneracy found experimentally in SS13 between the two models mimics an SPT, which thereby confirmed that it occurs in real lens modeling. 
To date, it is not clear whether the conclusions drawn in SS13 and SS14 about time delays and $H_0$ could be generalized to other mass distributions modified under exact SPTs or only 
reflect the behavior of a very special case. In this paper, we address this question by studying how time delays are sensitive to the effects of the SPT. 

This paper is organized as follows. 
For readers who are not familiar with the SPT, we outline its basic principles in Sect. \ref{section:SPTprinciple}. 
In particular, we recall the importance of identifying a validity criterion. 
Owing to the valuable insight it offers for more general cases, we consider the SPT-modified time delays for axisymmetric lenses in Sect. \ref{section:TDaxisymmetric}. 
For non-axisymmetric lenses, we discuss in Sect. \ref{section:TDnonaxisymmetric} the relevance of the validity criterion defined in USS17 and analyze the SPT-modified time delays 
in detail. We summarize our findings and conclude in Sect. \ref{section:conclusions}.

\section{The principle of the source position transformation}
\label{section:SPTprinciple}

This section focuses on the principle of the SPT and the most recent theoretical results. 
For a detailed discussion, the reader is referred to SS14 and USS17.
All the analytical results presented in this paper have been implemented into a user friendly 
\python\ package called \pythonpackage. All the numerical results and figures have also been obtained from \pythonpackage\ without using any extra software. For an overall description of the package,
we refer the reader to the companion paper \citet[][submitted]{pySPT_WO_ArXiv}.

The basic idea underlying the SPT can be simply summarized as follows. A given general mass distribution $\kappa(\bt)$ defines a deflection law $\ba(\bt)$ which
describes how the light paths are affected in the vicinity of the deflector. The $n$ lensed image angular positions $\bt_{i}$ of a point-like source at unobservable position $\bb$ are those 
which satisfy the lens equation $\bb = \bt_{i} - \ba(\bt_{i})$. 
Then, from astrometric observations we can infer the constraints 
\begin{equation}
	\bt_i - \ba(\bt_i) = \bt_j - \ba(\bt_j) \ ,
	\label{spt_constraints}
\end{equation}
for all $1 \leq i < j \leq n$, leading to the mapping $\bt_{i}(\bt_{1})$ defined by the relative image positions of the same source.
The SPT addresses the following question: can we define an alternative deflection law, denoted as $\bha(\bt)$, which preserves the mapping $\bt_{i}(\bt_{1})$ for a unique source? 
If such a deflection law exists, it will necessarily be associated with the alternative source position $\bhb = \bt_{i} - \bha(\bt_{i})$, defining a new lens mapping, in such 
a way that
\begin{equation}
	\bt = \bb + \ba(\bt) = \bhb + \bha(\bt) \ .
	\label{spt_implicit}
\end{equation}
An SPT consists in a global transformation of the source plane formally defined by a 
mapping $\bhb(\bb)$ 
which gives rise to the transformed deflection law
\begin{equation}
	\bha(\bt) = \ba(\bt) + \bb - \bhb(\bb) = \ba(\bt) + \bb - \bhb(\bt - \ba(\bt)) \ , 
	\label{hat_alpha_definition}
\end{equation}
where in the first step we used Eq.\,\eqref{spt_implicit} and in the last step we inserted the original lens equation.
The mapping $\bhb(\bb)$ is chosen so that it satisfies $\text{det} (\partial \bhb / \partial \bb) \neq 0$ for all $\bb$ in the region of interest, hence $\bhb(\bb)$ is one-to-one.
This property of the source mapping guarantees the pairing of images to be conserved.
With $\bha$ defined this way, Eq.\,\eqref{spt_implicit} guarantees that all images of a given source $\bb$ under the original deflection law $\ba(\bt)$ are also images of the 
source $\bhb$ under the modified deflection law $\bha(\bt)$. Therefore, the mapping $\bt_{i}(\bt_{1})$ is preserved for all source positions. 

From the Jacobi matrix $\hA(\bt) = \partial \bhb / \partial \bt = (\partial \bhb / \partial \bb)(\partial \bb / \partial \bt)$ of the modified lens mapping $\bhb = \bt - \bha(\bt)$, SS14 have shown that both the magnification ratios of image pairs and their relative shapes remain unchanged under an SPT. In general, the Jacobi matrix $\hA(\bt)$ is not symmetric, which indicates that the modified deflection law $\bha$ is not a curl-free field,  
\begin{equation}
	|\boldsymbol{\nabla} \times \bha(\bt)| = \left|\hat{\mathcal{A}}_{12}(\bt) - \hat{\mathcal{A}}_{21}(\bt)\right| \neq 0 \ ,
	\label{not_curl_free}
\end{equation}
where the subscript indices refer to the matrix entries.
Therefore, $\bha$ cannot be in general expressed as the gradient of a deflection potential $\hpsi$ and does not correspond to the deflection produced by a gravitational lens.
Thus, there exists no physical mass distribution $\hkp$ leading to the modified deflection law $\bha$. 
The only cases for which $\hA(\bt)$ is globally symmetric occur either when the SPT simply reduces to an MST $\bhb(\bb) = \lambda\,\bb$, 
or when axisymmetric lenses are transformed under SPTs corresponding to a general radial stretching of the form 
\begin{equation}
	\bhb(\bb) = \left[1 + f(|\bb|) \right] \bb \ ,
	\label{radial_stretching}
\end{equation} 
where $f$ is called the deformation function. 
For such cases, we can always define $\hkp$ so that $2\,\hkp = \bn \cdot \bha = \nabla^2 \hpsi$.
However, even in this case there is still no guarantee that $\hkp$ corresponds to a physical mass distribution. 
Depending on the SPT, the modified mass profile may become non-monotic or even non-positive definite in particular regions of the lens plane.

Provided the curl component of $\bha$ is sufficiently small, it was shown in USS17 that one can define a curl-free deflection law $\bta$
which is very similar to $\bha$ in the sense that their difference is smaller than the astrometric accuracy $\sub{\varepsilon}{acc}$ of current observations
\begin{equation}
	 |\bta(\bt) - \bha(\bt)| \eqqcolon |\Delta \ba(\bt)| < \sub{\varepsilon}{acc} \ ,
	\label{criterion}
\end{equation}
in a finite region
$\mathcal{U}$ where multiple images occur. Therefore, $\bta$ can be derived as the gradient of a deflection potential $\tpsi$, 
which is caused by a mass distribution $\tkp$ corresponding to a gravitational lens.
The central question of the validity of an SPT was addressed in USS17. Whereas $\bha$ yields exactly the same lensed image positions as the original lens, $\bta$ does not.  
Because of observational uncertainties and additional physical reasons such as substructures in the mass distribution, we cannot reproduce observed positions to better than a few milliarcseconds (mas) with 
a smooth mass model (for a detailed discussion see SS14). A given SPT $\bhb(\bb)$ 
should be flagged as being valid as long as a corresponding curl-free $\bta$ leads to lensed image shifts smaller than a few mas. 
In this sense, the SPT is only an approximate invariance transformation.
The condition \eqref{criterion} was chosen in USS17 as the criterion to assess whether an 
SPT is valid or not. The relevance of this choice is reconsidered in detail in Sect. \ref{subsection:validity}. 

Because it will be of practical interest for deriving SPT-modified time delays in the non-axisymmetric case (see Sect. \ref{section:TDnonaxisymmetric}), we recall 
here the explicit expressions for $\tpsi$ and $\bta = \bn \tpsi$. These can essentially be obtained by formulating the `action' 
\begin{equation}
	S = \int_{\mathcal{U}} \left| \bn \tpsi - \bha\right|^2 \text{d}^2\theta
	\label{action_unruh}
\end{equation}  
for which finding a minimum leads to the Neumann problem
\begin{eqnarray}
  \left\{
      \begin{aligned}
        &\nabla^2 \tpsi =  \bn \cdot \bha \eqqcolon 2 \hkp &\text{for all}\ &\bt \in \mathcal{U} \ ,\\
        &\bn \tpsi \cdot \boldsymbol{n} = \bha \cdot \boldsymbol{n} &\text{for all}\ &\bt \in \partial\mathcal{U} \ , 
      \end{aligned}
    \right.
    \label{Neumann}
\end{eqnarray}
where $\partial\mathcal{U}$ represents the boundary curve of $\mathcal{U}$ and $\boldsymbol{n}$ the outward directed normal vector. 
From Eq.\,\eqref{action_unruh}, we see that the condition \eqref{criterion} plays a central role in defining a curl-free counterpart $\bta$ of the SPT-modified deflection law $\bha$.
We also note that the first relation in Eq.\,\eqref{Neumann} implies $\tkp = \hkp$ for all $\bt \in \mathcal{U}$. 
The Neumann problem can be solved by means of a Green's function for which an analytical solution is known when $\mathcal{U}$ is a disk of radius $R$. 
Thus, the deflection potential $\tpsi$ evaluated at the position $\bvt$ in the lens plane explicitly reads \citep[][submitted]{SPT_USS17, pySPT_WO_ArXiv}
\begin{equation}
	\tpsi(\bvt) = \left\langle \tpsi \right\rangle + 2 \int_{\mathcal{U}} H_1(\bvt;\bt)\ \hat{\kp}(\bt)\ \text{d}^2\theta - \int_{\partial\mathcal{U}} H_2(\bvt;\bt)\ \bha \cdot \boldsymbol{n}\ \text{d}s \ ,
	\label{tilde_psi_simplified}
\end{equation}
where $\left\langle \tpsi \right\rangle$ is the average of $\tpsi$ on $\mathcal{U}$, d$s$ the line element of the boundary curve $\partial\mathcal{U}$, 
\begin{equation}
	H_1(\bvt;\bt) = \frac{1}{4 \pi} \left[\ln\left(\frac{\left|\bvt-\bt\right|^2}{R^2}\right) + \ln\left(1 - \frac{2 \bvt \cdot \bt}{R^2} + \frac{|\bvt|^2 |\bt|^2}{R^4}\right) - \frac{|\bt|^2}{R^2} \right] \ ,
	\label{H1}
\end{equation}
and
\begin{equation}
	H_2(\bvt;\bt) = \frac{1}{4 \pi} \left[2 \ln\left(\frac{\left|\bvt-\bt\right|^2}{R^2}\right) - 1 \right] \ .
	\label{H2}
\end{equation}
The corresponding deflection angle $\bta$ can be derived by 
obtaining the gradient of $H_1$ and $H_2$ with respect to $\bvt$, which reads 
\begin{eqnarray}
	\bta(\bvt) &=& \frac{1}{\pi} \int_{\mathcal{U}} \left(\frac{\bvt - \bt}{|\bvt - \bt|^2} + \frac{|\bt|^2 \bvt - R^2 \bt}{R^4 - 2 R^2 \bvt \cdot \bt + |\bvt|^2 |\bt|^2}\right)\ \hat{\kp}(\bt)\ \text{d}\bt \nonumber \\
	             &-& \frac{1}{\pi} \int_{\partial\mathcal{U}}\frac{\bvt - \bt}{|\bvt - \bt|^2} \ \bha \cdot \boldsymbol{n}\ \text{d}s \ .
	\label{tilde_alpha_simplified}
\end{eqnarray}
Of course, $\tpsi$ and $\bta$ depend on the radius $R$ of the circular region $\mathcal{U}$ and must satisfies the two conditions $R > |\bvt|$ and $R$ not too large to ensure 
the criterion \eqref{criterion} to be satisfied.

To quantitatively study the impact of the SPT on time delays, it will be necessary to explicitly define a mapping $\bhb(\bb)$. We will 
focus most of this work on an isotropic SPT described by the radial stretching of the form \eqref{radial_stretching}. 
In particular, we will consider the special case where the deformation function $f(|\bb|)$ is the lowest-order expansion of more 
general functions
\begin{equation}
	f(|\bb|) = f_0 + \frac{f_2}{2 \tE^2} |\bb|^2 \ ,
	\label{deformation_function}
\end{equation}
where $f_0 \coloneqq f(0)$, $f_2 \coloneqq \tE^2\,f''(0)$ and $\tE$ is the Einstein angular radius. When $f_2 = 0$, Eq.\,\eqref{deformation_function} reduces to $f_0$ and 
the radial stretching \eqref{radial_stretching} simplifies to a pure MST with $\lambda = 1 + f_0$.
Such as in SS14 and USS17, we only consider SPT parameters which yield to physically meaningful modified mass profiles.

\section{Time delays: the axisymmetric case}
\label{section:TDaxisymmetric}

Owing to its simplicity, the study of how an SPT affects time delays between lensed images produced by an axisymmetric lens 
provides a valuable insight on the general non-axisymmetric case. 
Since $\ba$ and $\bt$ are collinear, the original lens mapping 
becomes one-dimensional and reads $\beta = \theta - \alpha(\theta)$. We set $\beta > 0$ and only consider the two outer\footnote{The inner lensed image is most of the time not observed.} lensed images $\tA$ 
and $\tB$ located on opposite sides of the lens center, i.e., $\tB < 0 < |\tB| < \tA$. 
From Eq.\,\eqref{oTau}, we readily deduce the one-dimensional form of the original time delay $\Delta \sub{t}{AB}$ between the image pair $(\tA, \tB)$
\begin{equation}
	\Delta t_{\text{\scalebox{.9}{AB}}} =  \frac{D_{\Delta t}}{c} \Big[\tau(\tA) - \tau(\tB)\Big] \eqqcolon \frac{D_{\Delta t}}{c}  \Delta \tau_{\text{\scalebox{.9}{AB}}} \ .
	\label{oTD1D}
\end{equation}

The one-dimensional  radial stretching \eqref{radial_stretching} simply reads 
\begin{equation}
	\hb(\beta) = [1 + f(\beta)]\ \beta \ ,
	\label{radial_stretching_1D}
\end{equation}
where $f(-\beta) = f(\beta)$ to preserve the symmetry. With no loss of generality, $1 + f(\beta) + \beta\,\text{d}f(\beta)/\text{d}\beta > 0$ assures the SPT to be one-to-one.
For the axisymmetric case, the SPT is an exact invariance transformation. Thus, the deflection law $\bha$ is a curl-free field, $\bta = \bha$, and there  
exists a deflection potential $\hpsi$ such as 
\begin{equation}
	\frac{\text{d}\hpsi(\theta)}{\text{d}\theta} = \ha(\theta) = \alpha(\theta) - f(\beta(\theta))\ \beta(\theta) \ ,
	\label{gradhpsi_ha}
\end{equation}
where in the second step we used the one-dimensional form of Eq.\,\eqref{hat_alpha_definition}. 
From Eq.\,\eqref{oTau}, we deduce that the SPT-modified extra light travel time $\hT$ reads
\begin{equation}
	\hT(\theta) = \frac{D_{\Delta t}}{c} \left[\frac{1}{2} \left(\theta - \hb[\beta(\theta)]\right)^2 - \hpsi(\theta) \right] \eqqcolon \frac{D_{\Delta t}}{c}\ \htau(\theta) \ .
	\label{hTau1D}
\end{equation}
From Eqs. \eqref{oTD} and \eqref{hTau1D}, the SPT-modified time delay between image pair $(\tA, \tB)$ of the same source 
thus becomes
\begin{equation}
	\Delta \hat{t}_{\text{\scalebox{.9}{AB}}} = \hT(\tA) - \hT(\tB) = \frac{D_{\Delta t}}{c} \Big( \htau(\tA) - \htau(\tB) \Big) \eqqcolon \frac{D_{\Delta t}}{c}  \Delta \htau_{\text{\scalebox{.9}{AB}}} \ .
	\label{hTD1D}
\end{equation}

With Eqs. \eqref{oTD1D} and \eqref{hTD1D}, we show in Sect. \ref{subsection:axisym:hatTD} that the time delay ratios $\Delta \sub{\hat{t}}{AB} / \Delta \sub{t}{AB} \ (\equiv \Delta \sub{\htau}{AB} / \Delta \sub{\tau}{AB})$  
can be highly simplified, revealing an elegant expression in terms of $\beta$ and $\hb(\beta)$, and valid for any axisymmetric lens and deformation function $f(\beta)$. 
We also propose an equivalent form of this relation in terms of the original and SPT-modified mean surface mass densities. 
We illustrate the analytical results with some examples in Sect. \ref{subsection:axisym:examples}.

\subsection{The SPT-modified time delays}
\label{subsection:axisym:hatTD}

After substituting the one-dimensional form of Eq.\,\eqref{spt_implicit} and Eq.\,\eqref{gradhpsi_ha} into Eq.\,\eqref{hTau1D}, 
the SPT-modified extra light travel time reads $\hT(\theta_i) = \htau(\theta_i)\ D_{\Delta t}/c$ with 
\begin{equation}
	\htau(\theta_i) = \frac{1}{2} \Big[\alpha(\theta_i) - f(\beta(\theta_i))\ \beta(\theta_i) \Big]^2 - \psi(\theta_i) + \int_{0}^{\theta_i} f(\beta(\theta))\ \beta(\theta)\ \text{d}\theta \ , 
	\label{hTau}
\end{equation}
up to a constant independent of $\theta$, keeping in mind that $\beta(\theta) = \theta - \alpha(\theta)$.
Because of $\beta(\tA) = \beta(\tB) \eqqcolon \varbeta$, we have $f(\beta(\tA)) = f(\beta(\tB)) = f(\varbeta)$, 
and the SPT-modified time delays between the images $\tA$ and $\tB$ is given by $\Delta \hat{t}_{\text{\scalebox{.9}{AB}}} = \Delta \htau_{\text{\scalebox{.9}{AB}}}\ D_{\Delta t}/c$ with
\begin{equation}
	\Delta \htau_{\text{\scalebox{.9}{AB}}} = \Delta \tau_{\text{\scalebox{.9}{AB}}}  - f(\varbeta)\,\varbeta\,(\tA - \tB) +  \int_{|\tB|}^{\tA} f(\beta(\theta))\ \beta(\theta)\ \text{d}\theta \ .
	\label{hTD}
\end{equation}
Due to the lens symmetry, the integral over $[\tB, |\tB|]$ does not contribute to $\Delta \htau_{\text{\scalebox{.9}{AB}}}$. 
With no loss of generality, we thus integrate from $|\tB|$ instead of $\tB$ in Eq.\,\eqref{hTD}. 
To go a step further, the difference $\Delta \tau_{\text{\scalebox{.9}{AB}}}$ between the original Fermat potentials  
can also be written as
\begin{eqnarray}
	\Delta \tau_{\text{\scalebox{.9}{AB}}} &=& - \varbeta\ (\tA - \tB) + \frac{1}{2} \left(\tA^2 - \tB^2\right) - \Big( \psi(\tA) - \psi(\tB) \Big) \ , \\
				   				&=& - \varbeta\ (\tA - \tB) +  \int_{|\tB|}^{\tA} \beta(\theta)\ \text{d}\theta \ ,
	\label{oTD2}
\end{eqnarray}
where in the first step we used the original lens equations $\alpha(\tA) = \tA - \varbeta$ and $\alpha(\tB) = \tB - \varbeta$, 
and in the last step we used $\text{d}\psi(\theta)/\text{d}\theta = \theta - \beta(\theta)$. 
Combining Eqs. \eqref{hTD} and \eqref{oTD2}, we then obtain from Eq.\,\eqref{hTD1D} the SPT-modified time delay
\begin{equation}
	\Delta \sub{\hat{t}}{AB} = \Delta \sub{t}{AB} \left[1 + f(\varbeta)\right] + \frac{D_{\Delta t}}{c}\ \eAB \ , 
	\label{hTDfinal}
\end{equation}
where we define $\sub{\varepsilon}{AB}$ as
\begin{equation}
	\eAB =  \int_{|\tB|}^{\tA} \beta(\theta) \left[f(\beta(\theta)) - f(\varbeta)\right]\ \text{d}\theta \ .
	\label{EpsilonAB}
\end{equation}
For the special case of a pure MST, the deformation function $f$ is independent of $\beta$, namely $f(\beta(\theta)) = f(\varbeta) \equiv \lambda - 1$ with $\lambda \in \mathbb{R}$. 
Therefore, $\sub{\varepsilon}{AB} = 0$ and we find $\Delta \sub{\hat{t}}{AB} = \lambda\ \Delta \sub{t}{AB}$ for all axisymmetric lenses, as expected. 
Considering the radial stretching \eqref{radial_stretching_1D} 
and a singular isothermal sphere (SIS) lens model, we show explicitly in Appendix \ref{appendix:epsilonAB_SIS} that $\sub{\varepsilon}{AB} = 0$ also holds for all image pairs $(\tA, \tB)$, i.e., for $0 \leq \beta < \tE$.
In fact, simple analytical arguments reveal that, in general, $\eAB$ remains very small compared to the other terms in Eq.\,\eqref{hTDfinal} and can be neglected.  
The demonstration is explained in detail in Appendix \ref{appendix:epsilonAB_general}.
As a result, the time delay ratios $\Delta \hat{t} / \Delta t$ given in Eq.\,\eqref{hTDfinal} can be simply approximated by
\begin{equation}
	\frac{\Delta \hat{t}}{\Delta t} \approx 1 + f(\beta) \equiv \frac{\hb(\beta)}{\beta} \ ,
	\label{hTDratio1}
\end{equation}
where we have dropped the subscript \scalebox{.9}{AB} keeping in mind that the equation holds only for time delay ratios between the same pair of lensed images corresponding to the source
$\hb$ and $\beta$.
For a given radial stretching, Eq.\,\eqref{hTDratio1} shows that the ratios between SPT-modified and original time delays scale basically like $\hb/\beta$, implying that they
depend explicitly on the deformation function $f(\beta)$, as it is the case for the MST. 

As written, Eq.\,\eqref{hTDratio1} misleadingly suggests that the time delay ratio is insensitive to the original lens profile $\kappa$. Consider two original radial mass 
profiles $\kappa^{(1)}$ and $\kappa^{(2)}$, which are not related under an SPT, and consider a source position $\beta$. We locate the corresponding pairs of brighter lensed images by 
$(\tA^{(1)}, \tB^{(1)})$ and $(\tA^{(2)}, \tB^{(2)})$. 
For a given deformation function $f(\beta)$, Eq.\,\eqref{hTDratio1} says that $\Delta \sub{\hat{t}}{AB}^{(1)} / \Delta \sub{t}{AB}^{(1)} \approx \Delta \sub{\hat{t}}{AB}^{(2)} / \Delta \sub{t}{AB}^{(2)}$, 
but the two time delay ratios are evaluated at two different pairs of positions which depend on the lens models, i.e., $\tA^{(1)} \neq \tA^{(2)}$ and $\tB^{(1)} \neq \tB^{(2)}$. 
When $\kappa^{(2)}$ corresponds to a modified version of $\kappa^{(1)}$ under the SPT $\hb(\beta) = [1 + g(\beta)]\ \beta$ (with $g(\beta)$ satisfying the conditions given after Eq.\,\ref{radial_stretching_1D}), 
we have $\tA^{(1)} = \tA^{(2)}$ and $\tB^{(1)} = \tB^{(2)}$. 
However, this case can be reduced to an original radial mass profile $\kappa^{(1)}$ deformed by an SPT that is defined as the composition of two other SPTs 
such as  $\hb(\beta) = [1 + h(\beta)]\ \beta$ with $h(\beta) = [1 + f(\beta)][1 + g(\beta)] - 1$. Thus, this leads to $\Delta \hat{t} / \Delta t \approx 1 + h(\beta)$, in agreement with Eq.\,\eqref{hTDratio1}.

The SPT-modified mass profile $\hkp$ of a radial profile $\kp$ is also radial (SS14). Therefore, time delays $\Delta \sub{t}{AB}$ and $\Delta \sub{\hat{t}}{AB}$ 
should depend only on the image positions and the corresponding surface mass densities in the annulus defined between the images. In particular for $\Delta \sub{t}{AB}$, the major contribution comes from the mean surface mass 
density $\sub{\langle\kappa\rangle}{AB}$ in the annulus $\left|\tB\right| < \theta < \tA$ \citep[][]{Gorenstein_Falco_Shapiro_1988, Kochanek_2002_TD, Kochanek_Saas-Fee}. We will show next that the time delay 
ratios \eqref{hTDratio1} can be expressed only in terms of $\sub{\langle \kappa \rangle}{AB}$ and the corresponding SPT-modified $\sub{\langle \hkp \rangle}{AB}$. 
First, we can easily show that
\begin{eqnarray}
	\sub{\langle \kp \rangle}{AB} 	&\coloneqq& 	\frac{2}{\tA^2 - \tB^2} \int_{|\tB|}^{\tA} \theta\ \kp(\theta)\ \text{d}\theta = \frac{m(\tA) - m(|\tB|)}{\tA^2 - \tB^2} \nonumber \\
							&=&  		1 - \frac{\varbeta}{\tA - |\tB|} \ ,
	\label{meankappa}
\end{eqnarray}
where in the last step we used $m(\theta) = \theta\ \alpha(\theta)$, $\alpha(\tA) = \tA - \varbeta$, and $\alpha(\tB) = \tB - \varbeta$.
Similarly, we can easily deduce that
\begin{equation}
	\sub{\langle \hkp \rangle}{AB} = 1 - \frac{\hb(\varbeta)}{\tA - |\tB|} \ , 
	\label{meanhatkappa}
\end{equation}
where we first used $\hat{m}(\theta) = \theta\ \ha(\theta)$ and Eq.\,\eqref{gradhpsi_ha}, then $\alpha(\tA) = \tA - \varbeta$ and $\alpha(\tB) = \tB - \varbeta$.
Combining Eqs. \eqref{hTDratio1} to \eqref{meanhatkappa}, we thus obtain for the time delay ratio in terms of mean surface mass densities
\begin{equation}
	\frac{\Delta \hat{t}}{\Delta t} \approx \frac{\hb(\beta)}{\beta} = \frac{1 - \langle \hkp \rangle}{1 - \langle \kp \rangle} \ ,
	\label{hTDratio2}
\end{equation}
where we have once again dropped the \scalebox{.9}{AB} keeping in mind that the mean surface mass densities are evaluated in the annulus defined by the inner and outer radii $|\tB|$ and $\tA$, respectively.
As expected, Eq.\,\eqref{hTDratio2} shows that the ratio between SPT-modified and original time delays depends essentially on mean surface mass densities in the annulus $\left|\tB\right| < \theta < \tA$.
Finally, we note that the second equality in Eq.\,\eqref{hTDratio2} is exact.

\subsection{Some illustrative examples}
\label{subsection:axisym:examples}

\begin{figure}
	\centering
	\resizebox{\hsize}{!}{\includegraphics{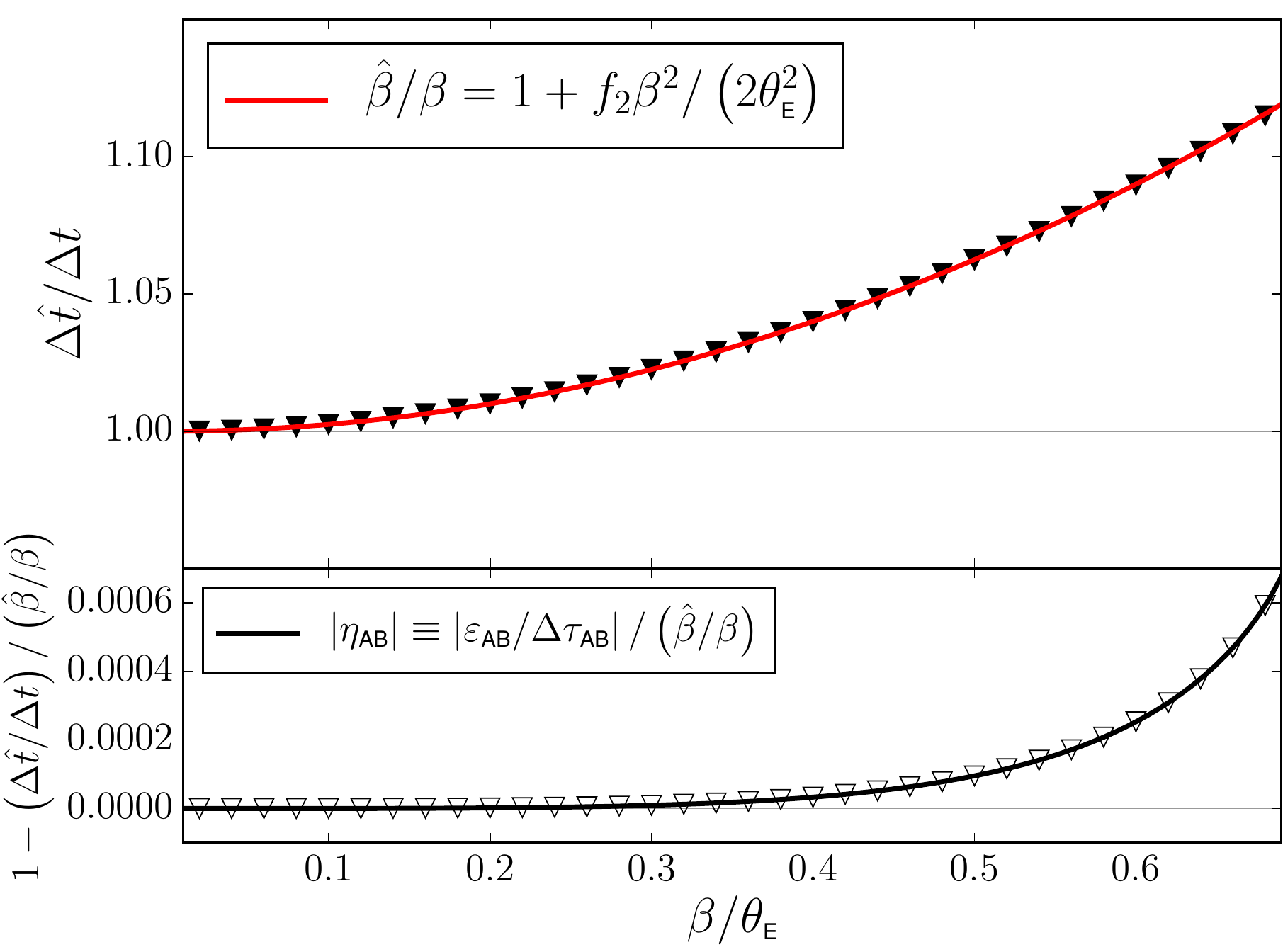}}
     	\caption{	Impact of an SPT described by the radial stretching $\hb(\beta) = 1 + f_2 \beta^2 / (2\,\theta_{\text{E}}^2)$, with $f_2 = 0.5$, on time delays image pairs generated 
			by a NIS, with core $\tc = 0.1\,\theta_{\text{E}}$. 
			\emph{Top:} ratio between SPT-modified and original time delays (black inverted triangles) for each source position. The time 
			delay ratios scale almost perfectly like $\hb/\beta$ (red curve), as predicted by Eq.\,\eqref{hTDratio1}. For a source close to the radial caustic $(\beta = 0.68\,\theta_{\text{E}})$, 
			the effect of the SPT reaches $\sim 11 \%$ and depends explicitly on the stretching parameter $f_2$. 
			\emph{Bottom}: numerical confirmation of the validity of the approximation performed in Eq.\,\eqref{hTDratio1} which consists in neglecting the term $\varepsilon_{\text{AB}}$. 
			The solid black curve illustrates $|\eta_{\text{AB}}|$ as an analytical function of $\beta$ and perfectly fits the quantity $1 - (\Delta \hat{t}/\Delta t) / (\hat{\beta}/\beta)$ numerically evaluated 
			for each source position. It appears clear that $|\eta_{\text{AB}}| \ll 1$ for all source positions $\beta$ which lead to multiple images.
		     }
       	\label{figure:DT_axisym_1}       
\end{figure}

To illustrate the results obtained in the previous section, we first consider the deformation function \eqref{deformation_function} with $f_2=0.5$ and $f_0=0$ to separate the
impact of the MST from that of the SPT. As original lens model, we choose a non-singular isothermal sphere (NIS) characterized by the deflection law
\begin{equation}
	\alpha(\theta) = \frac{\tE\,\theta}{\sqrt{\tc^2 + \theta^2}} \ ,
	\label{alpha_NIS}
\end{equation}
where the core radius $\tc$ is defined such as $\tc = \nu\,\tE$ with $0 < \nu < 1$.
To derive time delays in the axisymmetric case, we only consider the three lensed image configurations where the fainter central image is omitted. Thus, we need to sample the source positions inside the 
radial caustic of angular radius $\sub{\beta}{r} = \beta(\sub{\theta}{r})$ where $\sub{\theta}{r}$ represents the angular radius of the corresponding radial critical curve. 
For an NIS, $\sub{\beta}{r}$ is simply given by 
$\sub{\beta}{r} = \tE\,(1 - \nu^{2/3})^{3/2} \approx 0.695\,\tE$ for $\nu = 0.1$. 
Using this simple lens model and \pythonpackage, we create pairs of mock images for a uniform set of $34$ sources covering the range $\beta = 0.02\,\tE$ to $\beta = 0.68\,\tE < \sub{\beta}{r}$.

The top panel in Fig.\,\ref{figure:DT_axisym_1} shows $\Delta \hat{t} / \Delta t$ as a function of $\beta$ for the corresponding pairs of lensed images.
We see that the time delay ratios scale remarkably well like the function $1 + f(\beta) \equiv \hb(\beta)/\beta$, as predicted by Eq.\,\eqref{hTDratio1}. 
According to Eq.\,\eqref{hTDratio1}, the strongest effect of the SPT on time delays arises for a source as close as possible to the radial caustic, i.e., for $\beta \rightarrow \sub{\beta}{r}$.
Thus, in our first example, the theoretical maximum time delay ratio equals $\Delta \hat{t} / \Delta t \approx 1.12$ for $f_2 = 0.5$ and $\nu = 0.1$, leading to an impact of $12\%$ on $H_0$.
As long as it leads to a physical meaningful $\hkp$, larger (resp. smaller) values of $|f_2|$ lead to larger (resp. smaller) time delay ratios.
To quantitatively evaluate the accuracy of Eq.\,\eqref{hTDratio1}, we compare the numerically evaluated quantity $1 - (\Delta \hat{t}/\Delta t) / (\hat{\beta}/\beta)$ 
to unity, as shown in the bottom panel in Fig.\,\ref{figure:DT_axisym_1}.
From Eq.\,\eqref{hTDfinal}, it follows that $1 - (\Delta \hat{t}/\Delta t) / (\hat{\beta}/\beta) = |\nAB|$ 
with $\nAB \coloneqq (\sub{\varepsilon}{AB}/\Delta \sub{\tau}{AB})/(\hb/\varbeta) $.
The quantity $|\sub{\eta}{AB}|$ is smaller than $10^{-4}$ for $\beta \leq 0.5\,\tE$, reaching a maximum of $|\sub{\eta}{AB}| \approx 6 \times 10^{-4} \ll 1$ for $\beta = 0.68\,\tE$, confirming that $\sub{\varepsilon}{AB}$ 
can be neglected in Eq.\,\eqref{hTDratio1} in such a case. 
For an NIS deformed by a radial stretching characterized by Eq.\,\eqref{deformation_function}, it is possible to derive an analytical solution for $\sub{\varepsilon}{AB}$, hence for $|\sub{\eta}{AB}|$, 
by solving Eq.\,\eqref{EpsilonAB}. This analytical solution is represented 
in the bottom panel in Fig.\,\ref{figure:DT_axisym_1} and fits perfectly 
the numerical evaluations of $1 - (\Delta \hat{t}/\Delta t) / (\hat{\beta}/\beta)$ at each source position, as expected.

\begin{figure}
	\centering
   	\includegraphics[height=7cm]{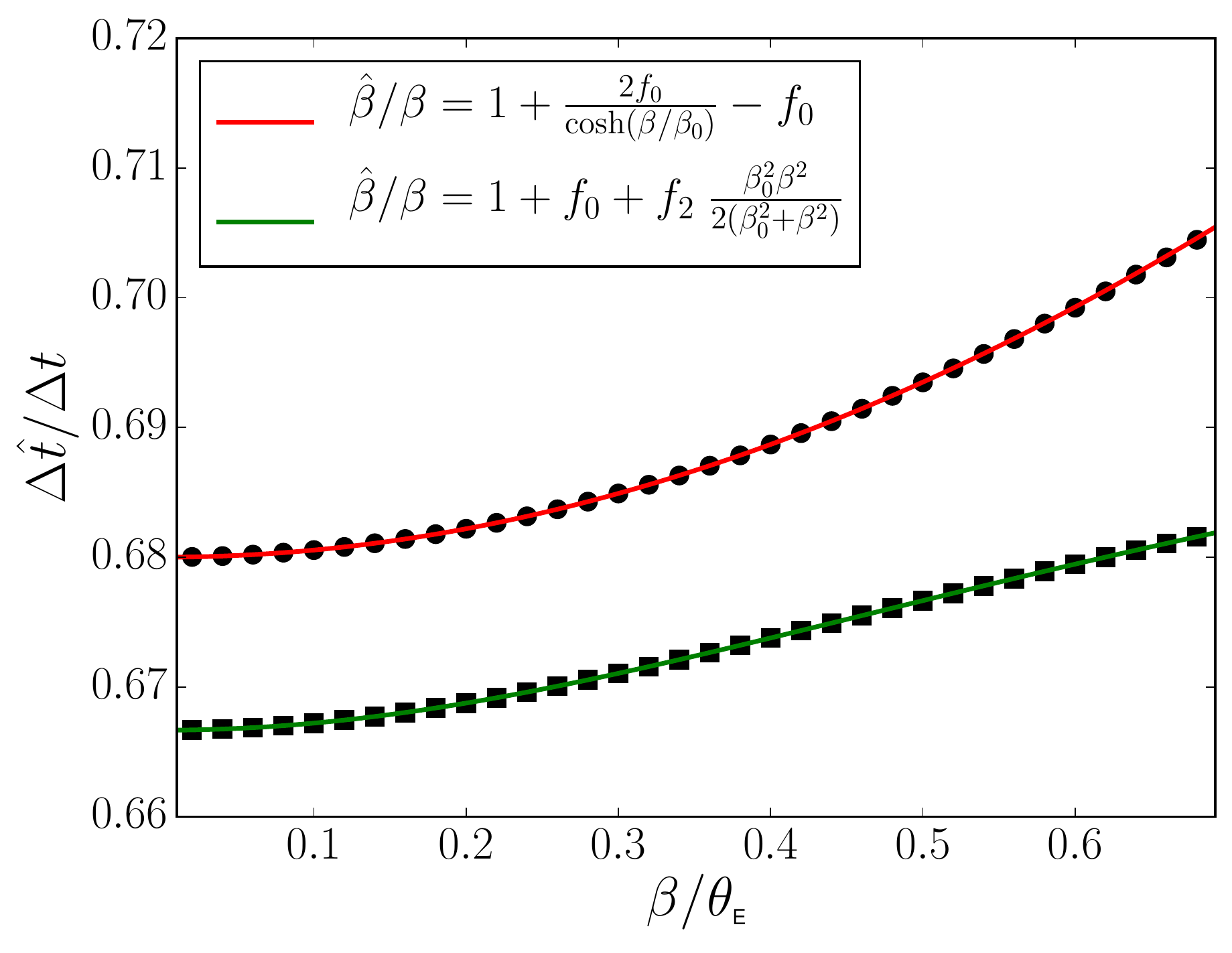}
     	\caption{	Impact on time delays of two different SPTs defined such that the corresponding $\hkp$ is approximately a power law near the tangential critical curve. 
			As predicted by Eq.\,\eqref{hTDratio1}, the time delay ratios (black dots and squares) scale almost perfectly like $\hb/\beta$ (red and green curves). 
			The ratio $\Delta \hat{t}/\Delta t$ for a pure MST is obtained when $\beta \rightarrow 0$, 
			which corresponds to $1 + f_0$ with $f_0 = -0.32$ and $f_0 = -1/3$, respectively. For a source 
			close to the radial caustic $(\beta = 0.68\,\theta_{\text{E}})$, the effect of the SPT on $H_0$ reaches around $3.6\%$ and 
			$2.2\%$.}
       	\label{figure:other_examples}       
\end{figure}

We have successfully tested the relation \eqref{hTDratio1} for various axisymmetric lens profiles deformed by different deformation functions. As additional examples, we consider the
two deformation functions
\begin{equation}
f(\beta) = \frac{2 f_0}{\cosh{(\beta / \beta_0)} - f_0} \ ,
\label{deffun_0}
\end{equation}
with $\beta_0 = \tE \sqrt{3 (1-f_0)/(1+f_0)}$ and $f_0 = -0.32$, and 
\begin{equation}
f(\beta) = f_0 + \beta_0^2\,f_2\,\beta^2 \ \left[2\,\left(\beta_0^2 + \beta^2\right)\right] \ ,
\label{deffun_1}
\end{equation}
with $\beta_0 = 0.8\,\tE$, $f_0 = - 1/3$ and $f_2 = 1/9$. 
The choice for the two deformation functions \eqref{deffun_0} and \eqref{deffun_1} is justified by the fact that the resulting SPT-modified mass profiles $\hkp$ are approximately power laws near the tangential 
critical curve, i.e., $\hkp(\theta) \approx \hkp(\tE) (\theta/\tE)^{-\upsilon}$ (SS14). In both cases, we adopt an NIS with $\tc = 0.1\,\tE$ as original lens model and the same source sample
as in the first example.
Fig.\,\ref{figure:other_examples} shows the time delay ratios between the lensed images for each source. As expected, $\Delta \hat{t} / \Delta t$ fits almost perfectly the function $1 + f(\beta) \equiv \hb(\beta)/\beta$.
For $\beta =  0$, the two deformation functions simplify to $f(\beta) = f_0$ and the corresponding SPTs reduce to pure MSTs leading to $\Delta \hat{t} / \Delta t = 0.68$ and $\Delta \hat{t} / \Delta t = 2/3$, respectively. 
Therefore, any changes from these values reflect the impact of the SPT. For $\beta = 0.68\,\tE$, Fig.\,\ref{figure:other_examples} shows an impact of around $3.6\%$ and $2.2\%$ on $H_0$, 
respectively, which is significantly smaller than what we have obtained for the first example.

Not all combinations of SPT deformation parameters and original mass profiles $\kappa$ yield a physically meaningful SPT-modified mass profile, namely $\hkp$ monotonically decreasing and positive definite \citep{SPT_SS14}. 
In addition, the maximum time delay ratio also depends on $\kappa$ since the latter directly defines the size of the radial caustic $\beta(\tr)$, namely the region in the source plane that 
produces multiple images. In summary, the way the SPT affects the time delays is very sensitive to the choice of the deformation function $f$, the associated deformation parameters, the original mass 
profile $\kappa$ and lensed image positions. For these reasons, we restrain ourselves to draw generalized quantitative conclusions 
in the axisymmetric case. However, our numerical tests suggest an effect of a few percent in general. We will show in the next section that the simple connection between the time delay ratios and the 
source position ratios may still be very strong in the non-axisymmetric case.

\section{Time delays: the non-axisymmetric case}
\label{section:TDnonaxisymmetric}

In this section, we drop the axisymmetry assumption for the original lens model. The SPT-modified deflection law $\bha$ is thus 
not a curl-free field in general and there exists no deflection potential $\hpsi$ which satisfies $\bn \hpsi = \bha$. To define a physically meaningful modified 
extra light travel time, we consider the deflection law $\bta$, the closest curl-free approximation to $\bha$ which satisfies the criterion \eqref{criterion}
for all $\bt$ over a region $\mathcal{U}$ (see Eq.\,\ref{tilde_alpha_simplified}), and the associated deflection potential $\tpsi$ (see Eq.\,\ref{tilde_psi_simplified}). 
For the rest of this section, we follow USS17 and condider
$\sub{\varepsilon}{acc} \approx 5 \times 10^{-3}\,\tE$ over the circular region $|\bt| \leq 2\,\tE$, where 
the approximation stems from the typical positional accuracy of the Hubble space telescope.

Within the region $\mathcal{U}$, the lensed images $\btt$ of the source $\bhb$ satisfying the lens mapping $\bhb = \btt - \bta(\btt)$ are expected to be 
sufficiently close to the corresponding original images $\bt = \bb + \ba(\bt) = \bhb + \bha(\bt)$ to not be distinguished observationally.
However, we show in Sect. \ref{subsection:validity} that the criterion \eqref{criterion} defined in USS17 
cannot guarantee the difference $|\Delta \bt| \coloneqq |\btt - \bt|$ between the SPT-modified image position $\btt$ of the source $\bhb$ and the image position $\bt$ of the source $\bb$  
to be smaller than $\sub{\varepsilon}{acc}$ over the whole region $\mathcal{U}$.
However, for specific pairs of original and SPT-modified sources leading to indistinguishable image configurations, we illustrate in Sect. \ref{subsection:timedelaysforvalid} the 
typical behavior of the time delay ratios. 
Finally, based solely on a numerical optimization, we slightly modify the source mapping $\bhb(\bb)$ by relaxing the isotropic condition of the SPT. It follows that the region where 
$|\Delta \bt| < \sub{\varepsilon}{acc}$ can be substantially extended.
From this ad hoc source mapping, $\bta$ and $\tpsi$, we illustrate the corresponding alternative time delay ratios in Sect. \ref{subsection:alternativetimedelays}.

\subsection{Criterion for the validity of an SPT}
\label{subsection:validity}

\begin{figure}
	\centering
   	\includegraphics[height=7cm]{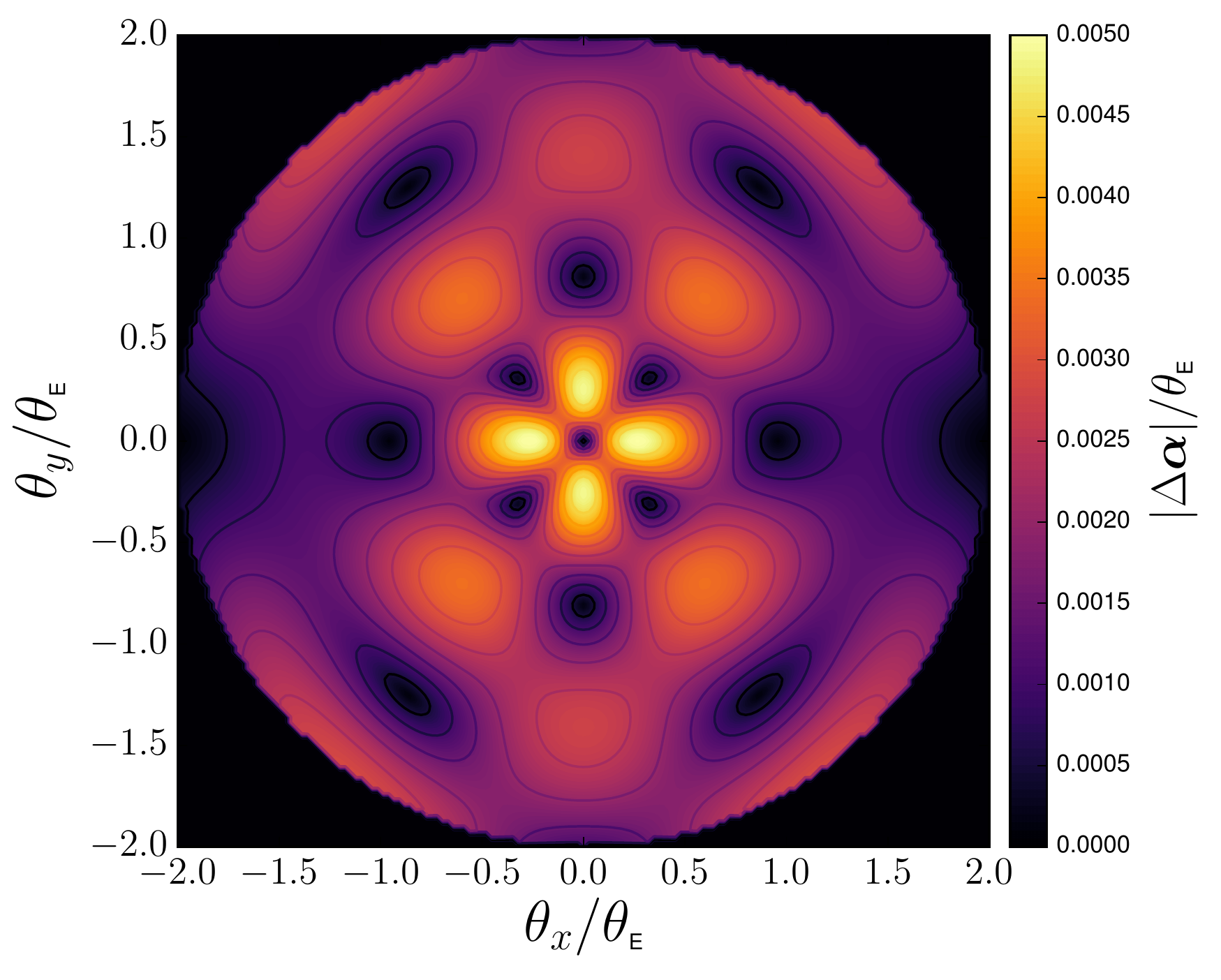}
     	\caption{	Map of $|\Delta \ba(\bt)|$ over the circular region $|\bt| \leq 2\,\theta_{\text{E}}$ for $f_2 = 0.4$, $\theta_{\text{c}} = 0.1\,\theta_{\text{E}}$ and $\gamma_{\text{p}} = 0.1$. 
			This figure is similar to the figure $7$ in \cite{SPT_USS17} with $f_2 = 0.55$, even though it is based on a different approach (see the text for more details). 
			The reason is that the relative values of $|\Delta \ba(\bt)|$ do not depend on $f_2$ but on $R$ and $\gamma_{\text{p}}$. 
			While $R$ defines the radius of the integration area $\mathcal{U}$ in Eq.\,\eqref{tilde_alpha_simplified}, the shear amplitude $\gamma_{\text{p}}$ 
			is the unique parameter which explicitly characterizes the degree of asymmetry of the original lens model.
			}
       	\label{figure:Delta_alpha_map_1}       
\end{figure}

\begin{figure}
	\centering
   	\includegraphics[width=9.2cm]{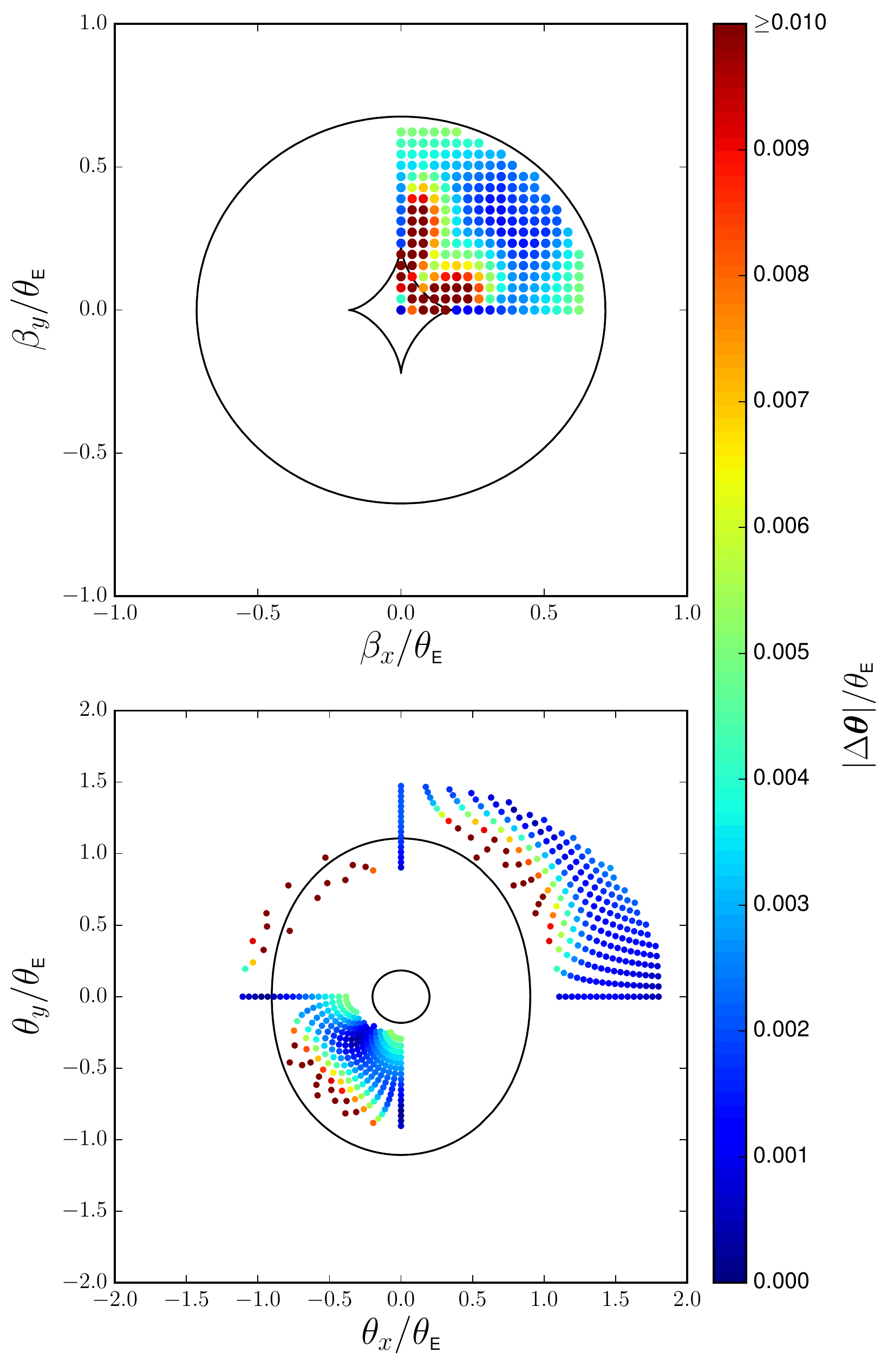}
     	\caption{	\emph{Top}: 	grid of source positions $\bb$ covering the radial range $0 \leq |\bb| \leq 0.66\,\theta_{\text{E}}$ in the first quadrant of the source plane. Each source produces a set of lensed images $\bt$ 
						(shown in the \emph{bottom panel}) under the original deflection law $\ba$ and the corresponding set of $\btt$ under the curl-free deflection law $\bta$. The color-coding refers to the largest offset 
						$|\Delta \bt|_{\text{max}}$ associated with each source. The solid black curves locate the caustic curves (\emph{top panel}) and the critical curves (\emph{bottom panel}), respectively, for the 
						NIS plus external shear.
			\emph{Bottom}:	set of mock lensed images $\bt$ produced by the source positions shown in the \emph{top panel} and lensed by an NIS plus external shear characterized by $\tc = 0.1\,\theta_{\text{E}}$ and 
						$\gamma_{\text{p}} = 0.1$. The color-coding represents the offsets $|\Delta \bt|$ in units of $\theta_{\text{E}}$ between $\bt$ and the images $\btt$. The latter are the images of the source 
						positions $\bhb$ lensed 
						by the SPT-modified lens associated with the curl-free deflection field $\bta$. Even though the criterion $|\Delta \ba(\bt)| < \varepsilon_{\text{acc}}$ is satisfied over 
						$|\bt| \leq 2\,\theta_{\text{E}}$, the offsets for most of the images located nearby the critical curves are larger than $\varepsilon_{\text{acc}}$. 
			}
       	\label{figure:tilde_images}       
\end{figure} 

\begin{figure}	
	\centering
	\includegraphics[height=7cm]{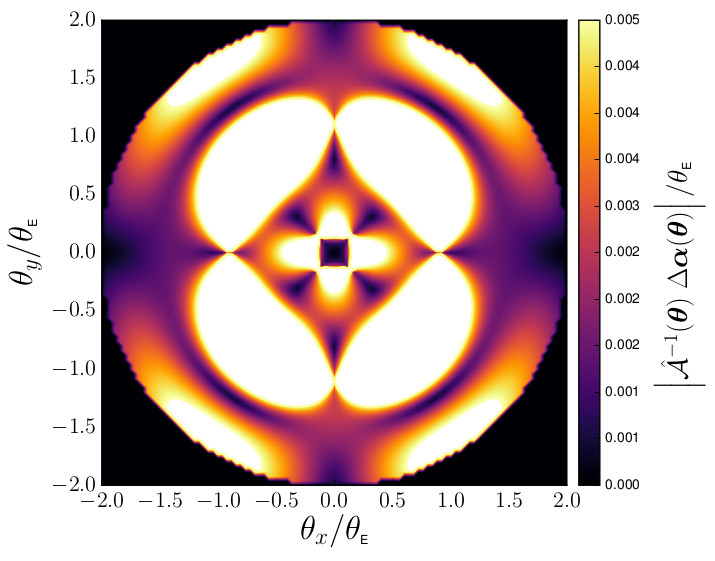}	
	\includegraphics[height=7cm]{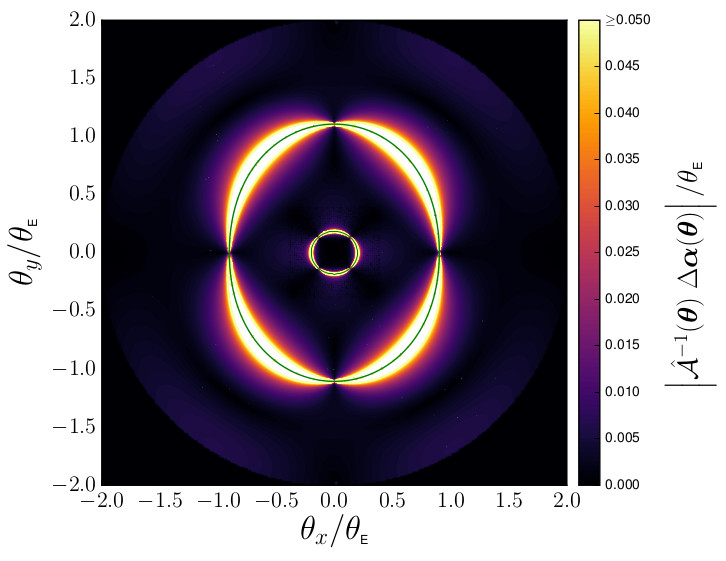}	
     	\caption{	Maps of $\left| \hat{\mathcal{A}}^{-1}(\bt)\ \Delta \ba(\bt)\right| \approx |\Delta \bt|$ in units of $\theta_{\text{E}}$ for $f_2 = 0.4$ and $\gamma_{\text{p}} = 0.1$. 
			\emph{Top}: 	we use the same color-coding as in Fig.\,\ref{figure:Delta_alpha_map_1} to explicitly show that $|\Delta \ba(\bt)| < 5 \times 10^{-3}\,\theta_{\text{E}}$ 
						over a region of the lens plane does not guarantee the image offsets $|\Delta \bt|$ to be smaller than $5 \times 10^{-3}\,\theta_{\text{E}}$ over the same 
						region.
			\emph{Bottom}:	we adjust the color-coding to bring out regions where the image offsets $|\Delta \bt|$ are the largest, namely the critical curves.
			}
			
       	\label{figure:Delta_theta_map_firstorder}       
\end{figure}

\begin{figure}
	\centering
	\includegraphics[height=7cm]{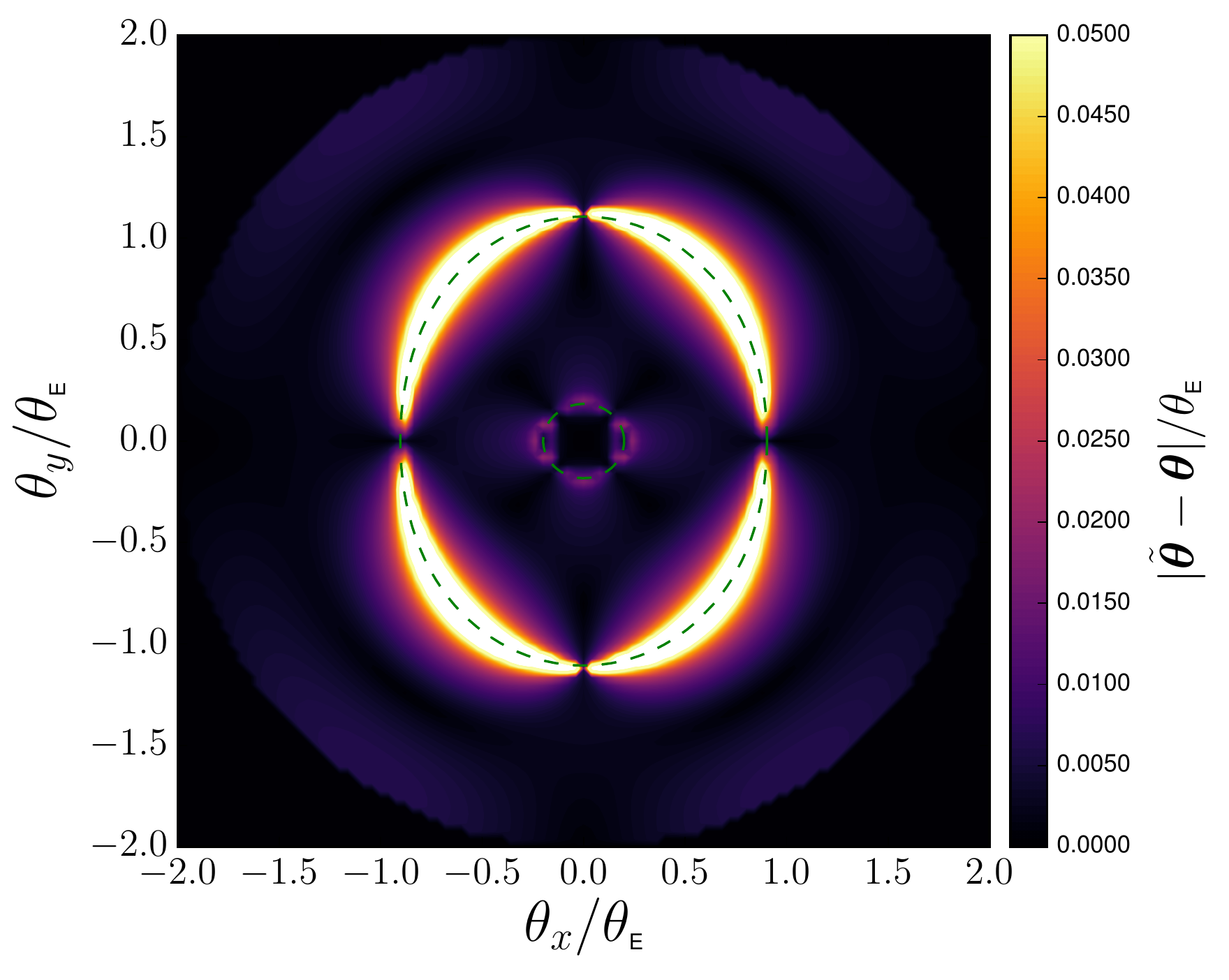}	
     	\caption{	Map of $\big|\btt - \bt\big| \eqqcolon |\Delta \bt|$ in units of $\theta_{\text{E}}$ for $f_2 = 0.4$ and $\gamma_{\text{p}} = 0.1$. Clear differences with Fig.\,\ref{figure:Delta_theta_map_firstorder}
			are observed for positions $\bt$ located almost on the radial critical curve, where the approximation adopted in Eq.\,\eqref{DeltaTheta} is expected to become not valid. This map is much more 
			time consuming to obtain than the ones represented in Fig.\,\ref{figure:Delta_theta_map_firstorder}.}
       	\label{figure:Delta_theta_map_images}       
\end{figure}

To illustrate the limit of the criterion $|\Delta \ba(\bt)| < \sub{\varepsilon}{acc}$, we consider a situation similar to SS14 and USS17, namely a quadrupole lens 
composed of an NIS plus external shear $\sub{\gamma}{p}$ (NISg) for which the deflection law is defined by
\begin{equation}
	\ba(\bt) = \frac{\tE\,\bt}{\sqrt{\tc^2 + |\bt|^2}} - \sub{\gamma}{p} \begin{pmatrix} 1 & 0 \\ 0 & -1 \end{pmatrix}\,\bt \ ,
	\label{quadrupole}
\end{equation}
where the core radius is set to $\tc = 0.1\,\tE$. 
The original source mapping is transformed by a radial stretching \eqref{radial_stretching} with a deformation function $f$ of the form \eqref{deformation_function}. The adopted SPT is thus defined by 
\begin{equation}
	\bhb(\bb) = \left(1 + f_0 + \frac{f_2}{2\,\tE^2} |\bb|^2\right) \bb \ .
	\label{radialstretch}
\end{equation}
For the rest of this section, we set the deformation parameter $f_2$ to be $f_2 = 0.4$ and the external shear magnitude to be $\sub{\gamma}{p} = 0.1$. 
Furthermore, we exclude the effect of a pure MST 
by simply choosing $f_0 = 0$. 
According to USS17 (see their figure 4), 
this specific pair $(f_2, \sub{\gamma}{p})$ constitutes an allowed pair of parameters in a sense it fulfills the criterion $|\Delta \ba(\bt)| < \sub{\varepsilon}{acc}$ over the circular 
region $|\bt| \leq 2\,\tE$.
Fig.\,\ref{figure:Delta_alpha_map_1} shows the map $|\Delta \ba(\bt)|$ over a circular grid $|\bt| \leq 2\,\tE$ in the lens plane. 
This figure is similar to the map $|\Delta \ba(\bt)|$ illustrated in the figure 7 in USS17, although they used $f_2 = 0.55$. 
It turns out that the ratio $|\Delta \ba(\bt_i)| / |\Delta \ba(\bt_j)|$ 
remains unaffected when $f_2$ varies, but is sensitive to variations of $R$ or $\sub{\gamma}{p}$. 
Actually, $\sub{\gamma}{p}$ is the only parameter that explicitly characterizes the degree of asymmetry of the original lens model.

The next step consists in determining how well the deflection law $\bta$ allows us to reproduce the original lensed images of a source. 
To this aim, we create a set of mock images $\bt$ of a sample of sources $\bb$ that cover the first quadrant of the source plane. We restrict the grid of sources
to $0 \leq |\bb| \leq 0.66\,\tE$ where multiple images are produced (see top panel in Fig.\,\ref{figure:tilde_images}). Then, we produce the images $\btt$ of the 
corresponding SPT-modified sources $\bhb(\bb)$.
The bottom panel in Fig.\,\ref{figure:tilde_images} shows the image positions $\bt$ and the color-coding represents 
$|\Delta \bt|$ in units of $\tE$. 
The same color-coding is applied to the sources (top panel) where only the largest offset, 
denoted by $|\Delta \bt|_{\text{max}}$, are shown.  

An `unexpected' conclusion can be drawn from Fig.\,\ref{figure:tilde_images}; even though  $|\Delta \ba(\bt)| < \sub{\varepsilon}{acc}$ over the region $|\bt| \leq 2\,\tE$ (as shown in Fig.\,\ref{figure:Delta_alpha_map_1}), 
many of image configurations are characterized by $|\Delta \bt| \gg \sub{\varepsilon}{acc}$ for at least one lensed image. This implies that these image configurations can be distinguished
from the original ones and the corresponding SPT can no longer be flagged as valid.
Furthermore, the largest offsets $|\Delta \bt|$ occur near 
the tangential critical curve. 
It comes with no surprise that the corresponding 
regions in the source plane are thus located near the tangential caustic curve.
To address this behavior, we first consider what the quantity $|\Delta \ba(\bt)|$ really represents. As defined in Eq.\,\eqref{criterion}, both $\bha$
and $\bta$ are evaluated at the same position $\bt$ in the lens plane. Therefore, we have $\bha(\bt) = \bt - \bhb$ and $\bta(\bt) = \bt - \btb$ where $\btb$ is 
the source position of the image $\bt$ under the deflection law $\bta$. 
Combining the two latter equations leads to
\begin{equation}
	\Delta \ba(\bt) = \bhb(\bt) - \btb(\bt) \eqqcolon \Delta \bb \ .
	\label{diffBeta}
\end{equation}  
Equation\,\eqref{diffBeta} shows that minimizing $|\Delta \ba(\bt)|$ is equivalent to minimizing $|\Delta \bb|$ with no guarantee on $|\Delta \bt|$.
Indeed, let us consider a position $\bt$ close to a critical line for which $|\Delta \ba(\bt)| < \sub{\varepsilon}{acc}$, for example $(\theta_x/\tE, \theta_y/\tE) = (0.5, 1.0)$ (see Fig.\,\ref{figure:Delta_alpha_map_1} and 
bottom panel in Fig.\,\ref{figure:tilde_images}). The corresponding source $\bb(\bt)$ is necessarily close to a caustic, so is $\bhb(\bt)$. Thus, the source $\btb(\bt)$ lies in a region of the source 
plane where even small shifts $|\Delta \bb|$ can lead to significantly different image positions. This explains why regions where $|\Delta \bt| \gg \sub{\varepsilon}{acc}$ are those which surround the critical curves.
Furthermore, whereas $\bt_i - \bha(\bt_i) = \bt_j - \bha(\bt_j)$ is satisfied for all $i \leq j$, 
we have $\bt_i - \bta(\bt_i) \neq \bt_j - \bta(\bt_j)$, meaning that the $\bt_i$ are not lensed images of a unique source under the deflection law $\bta$. 
Thus, the criterion $|\Delta \ba(\bt_{i})| < \sub{\varepsilon}{acc}$ for a lensed image configuration $\bt_{i}$ is based upon positions that are not linked under the deflection law $\bta$.
These few simple arguments suggest with no loss of generality that the choice of Eq.\,\eqref{criterion} as a validity criterion may not be the most appropriate one. 

Let us now evaluate $\bta$ at the position $\btt = \bhb + \bta(\btt)$ instead of $\bt$ and consider the difference $\bta(\btt) - \bha(\bt)$. We readily find that
\begin{equation}
	 \left|\bta\left(\btt\right) - \bha(\bt)\right| = \left|\btt - \bt \right| = \left|\Delta \bt \right|\ ,
	\label{diffAlpha_tilde}
\end{equation}  
which corresponds exactly to the image shift induced by the SPT that we expect to be smaller than $ \sub{\varepsilon}{acc}$. 
Assuming that $\left|\Delta \bt \right|$ is small, we can show to first order that
\begin{eqnarray}
	\bhb 	&=& \btt - \bta(\btt) \ ,\nonumber\\
		&=& \bt + \Delta \bt - \bha(\bt + \Delta \bt) - \Delta \ba (\bt + \Delta \bt) \ , \nonumber \\
		&\approx& \bt - \bha(\bt) + \left(1 - \frac{\partial \bha}{\partial \bt}\right)\,\Delta \bt - \Delta \ba(\bt) \ .
	\label{diffAlpha_tilde_firstorder}
\end{eqnarray}  
Thus, for all positions $|\bt| \leq 2\,\tE$ not located on a critical curve, Eq.\,\eqref{diffAlpha_tilde_firstorder} leads to
\begin{equation}
	|\Delta \bt|  \approx \left| \hat{\mathcal{A}}^{-1}(\bt)\ \Delta \ba(\bt)\right| \ .
	\label{DeltaTheta}
\end{equation} 
Equation\,\eqref{DeltaTheta} clearly shows that the offsets $|\Delta \bt|$ are related to $|\Delta \ba(\bt)|$ through 
the SPT-modified Jacobi matrix $\hat{\mathcal{A}}(\bt)$ whose impact 
become larger as we get closer to the critical curves. 
Figure\,\ref{figure:Delta_theta_map_firstorder} illustrates the quantity $\left| \hat{\mathcal{A}}^{-1}(\bt)\ \Delta \ba(\bt)\right|$ 
using two different color-coding. 
The upper panel shows the same color-coding as used in Fig.\,\ref{figure:Delta_alpha_map_1} for comparison. 
It shows that a significant part of the region $|\bt| < 2\,\tE$ is characterized by $|\Delta \bt| > \sub{\varepsilon}{acc}$ even though $|\Delta \ba(\bt)| < \sub{\varepsilon}{acc}$.
The lower panel adopts a color-coding which allows us to better visualize regions characterized by the largest offsets. These regions surround
the two critical curves represented by the two green lines. 
We confirm the validity of the first order Eq.\,\eqref{DeltaTheta} by comparing Fig.\,\ref{figure:Delta_theta_map_firstorder} with Fig.\,\ref{figure:Delta_theta_map_images}, which represents explicitly the quantity 
$\left|\btt - \bt\right|$. As expected, small differences can be observed very close to the critical curves where higher order terms in Eq.\,\eqref{diffAlpha_tilde_firstorder} become significant and cannot be ignored.
In addition, Fig.\,\ref{figure:Delta_theta_map_images} is much more time consuming to obtain than Fig.\,\ref{figure:Delta_theta_map_firstorder}. 
Indeed, a single $\left|\btt - \bt\right|$ evaluation requires $\btt$ to be calculated first, that is solving the lens equation $\bhb = \btt - \bta(\btt)$ that implies numerous $\bta$ evaluations.
In contrast, a single Eq.\,\eqref{DeltaTheta} evaluation requires only one $\bta$ evaluation. For this reason, the grid density in bottom panel in Fig.\,\ref{figure:Delta_theta_map_firstorder} is $20$ times 
higher than in Fig.\,\ref{figure:Delta_theta_map_images}.

Eq.\,\eqref{DeltaTheta} confirms that the criterion $|\Delta \ba(\bt)| < \sub{\varepsilon}{acc}$ for the validity of an SPT cannot guarantee the angular separation $|\Delta \bt|$ to be smaller than the astrometric 
accuracy of current observations, at least in regions nearby critical curves.
To construct the curl-free deflection field $\bta$, USS17 have considered the `action' defined in Eq.\,\eqref{action_unruh}
for which they found a minimum. This approach is based explicitly on the validity criterion \eqref{criterion}, 
which is not satisfactory 
and should be reconsidered. 
A new appropriate criterion would of course imply the definition of a new `action' to be minimized, leading to a new definition for $\bta$ and $\tpsi$.
Such a new approach is beyond the scope of this paper and will not be addressed here. 
Nevertheless, it remains possible to quantitatively estimate the impact of the SPT 
on time delays with the means available. In the next section, we first focus on the subset of source positions $\bb$ depicted in the top panel in Fig.\,\ref{figure:tilde_images}
that yields $|\Delta \bt| < \sub{\varepsilon}{acc}$.

\subsection{The SPT-modified time delays for valid configurations}
\label{subsection:timedelaysforvalid}

\begin{figure}
	\centering
	\includegraphics[height=7cm]{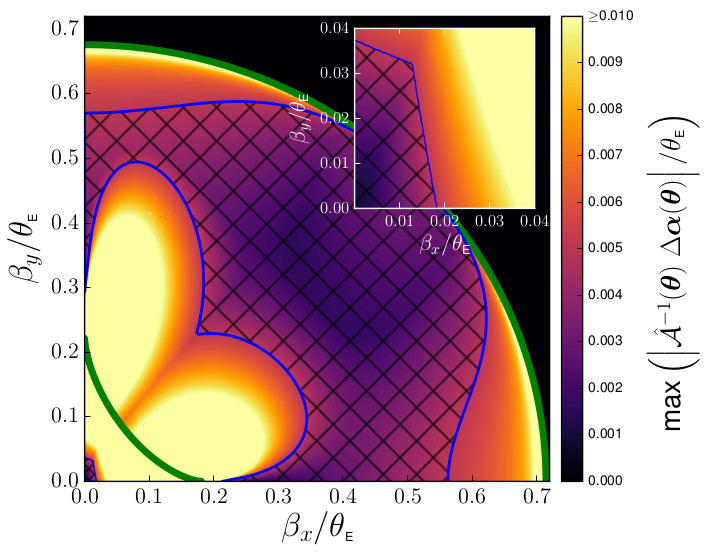}
     	\caption{	Map of $|\Delta \bt(\bb)|_{\text{max}}$ in units of $\theta_{\text{E}}$ for $f_2 = 0.4$ and $\gamma_{\text{p}} = 0.1$ (see Eq.\,\ref{SPregion_feature}). 
			The inner (resp. outer) green line represents the tangential (resp. radial) caustic curve. The two hatched regions ($\mathit{B}_1$ and $\mathit{B}_2$) delimited by blue curves demarcate parts of 
			the source plane where $|\Delta \bt(\bb)|_{\text{max}} \leq 5 \times 10^{-3}\,\theta_{\text{E}}$. The region $\mathit{B}_1$ lies inside the tangential caustic curve while $\mathit{B}_2$ lies outside. 
			The inset highlight the region around the position $\bb = \boldsymbol{0}$.
		     }
       	\label{figure:Delta_theta_map_sourceplane}       
\end{figure}

\begin{figure}
	\centering
	\includegraphics[height=7cm]{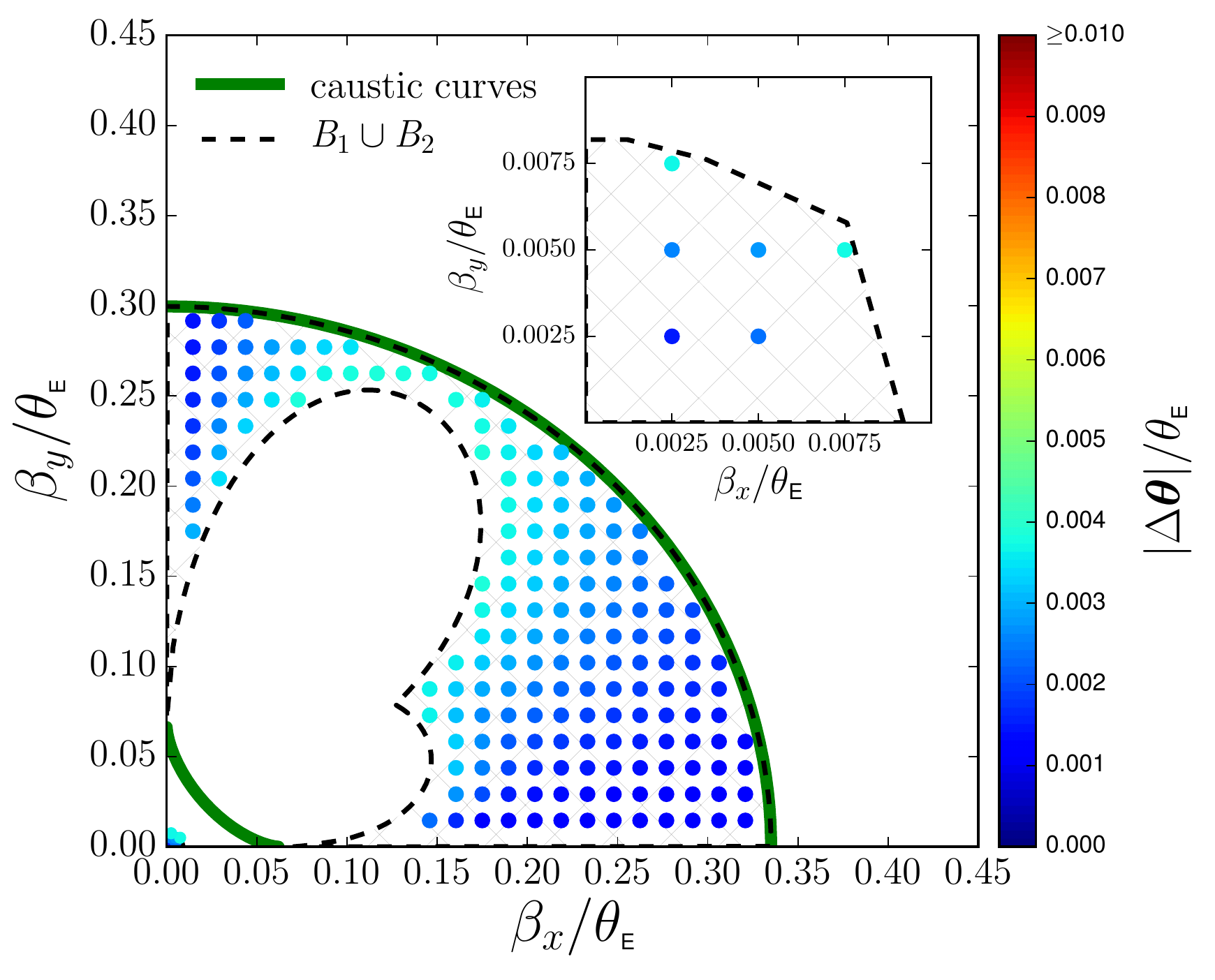}
     	\caption{	Grid of sources $\bb$ located inside the region $\mathit{B}_1 \cup \mathit{B}_2$ for the NIE with $(\theta_{\text{c}}, \epsilon) = (0.1\,\theta_{\text{E}}, 0.15)$. 
			The inner (resp. outer) green line represents the tangential (resp. radial) caustic curve. The color-coding refers to the offsets $|\Delta \bt|/\theta_{\text{E}}$ between
			the lensed images $\bt$ of the sources $\bb$ and the lensed images $\btt$ of the SPT-modified sources $\bhb$.}
       	\label{figure:Delta_theta_map_sourceplane_NIE}       
\end{figure}
 
In the previous section, we have shown that the criterion defined in Eq.\,\eqref{criterion} does not guarantee $|\Delta \bt| < \sub{\varepsilon}{acc}$ for all $|\bt| \leq 2\,\tE$. However, the top panel in 
Fig.\,\ref{figure:tilde_images} also shows sources (mainly outside the tangential caustic curve) for which the corresponding largest offsets  $|\Delta \bt|_{\text{max}}$ between 
original and SPT-modified image configurations are smaller than $\sub{\varepsilon}{acc}$.
Adopting the same original lens model as in the previous section, Fig.\,\ref{figure:Delta_theta_map_sourceplane} shows the quantity 
\begin{equation}
	|\Delta \bt(\bb)|_{\text{max}} \coloneqq \text{max}\left( \left| \hat{\mathcal{A}}^{-1}(\bt(\bb))\ \Delta \ba(\bt(\bb))\right|\right) 
	\label{SPregion_feature}
\end{equation} 
in units of $\tE$ over the first quadrant in the source plane. The region outside the radial caustic curve 
is irrelevant in our case 
because it does not lead to multiple image configurations.
The blue lines demarcate two disjointed hatched regions, denoted by $\mathit{B}_1$ and $\mathit{B}_2$, so that all sources inside $\mathit{B}_1 \cup \mathit{B}_2$ lead 
to image configurations characterized by $|\Delta \bt|_{\text{max}} \leq \sub{\varepsilon}{acc}$. For $f_2 = 0.4$ and $\gamma_{\text{p}} = 0.1$, the region $\mathit{B}_1 \cup \mathit{B}_2$ 
covers around $56\%$ of the area enclosed by the radial caustic curve. Smaller values for $f_2$ yield larger $\mathit{B}_1 \cup \mathit{B}_2$ areas, up to $100\%$ when $f_2 = 0$ 
(SPT reduced to an MST) or $\gp = 0$ (axisymmetric lens). The regions $\mathit{B}_1$ and $\mathit{B}_2$ are situated on both sides of the tangential caustic curve. 
The very high area ratio between these two regions ($1$ to $560$ in this case) indicates that most of the valid image configurations are composed of two images (the fainter 
third central one is always omitted). Moreover, the few `valid' four component configurations are very symmetric, suggesting comparable time delays between opposite image pairs. 

Provided that $\hkp$ is physically meaningful, the curl-free deflection field $\bta$ yields indistinguishable image configurations for sources $\bb \in \mathit{B}_1 \cup \mathit{B}_2$ 
as compared to the original $\ba$. Although these valid image configurations are of limited interest for time delay cosmography\footnote{The sources located inside the region $\mathit{B}_1$ produce very symmetric 
quadruply imaged configurations, while those located inside the region $\mathit{B}_2$ produce only doubled image configurations. In both cases, only one relevant time delay can be inferred from these systems.}, 
the resulting model ambiguities may still prevent us
from performing a robust lens modeling. Thus, even though the adopted SPT is not `valid' over all the region $|\bt| \leq 2\,\tE$, we propose in this section to analyze the time delay ratios 
of these particular image configurations between the original and SPT-modified models.
To this aim, we consider an original non-axisymmetric mass distribution which produces $n$ lensed images $\bt_i$ of a source $\bb \in \mathit{B}_1 \cup \mathit{B}_2$. The time delay $\Delta t_{ij}$ between a 
pair of lensed images $\bt_i$ and $\bt_j$
is defined in Eq.\,\eqref{oTD}. The corresponding SPT-modified time delay $\Delta \tilde{t}_{ij}$ have to be evaluated at image positions $\btt_i$ and $\btt_j$, respectively, leading to 
\begin{equation}
	\Delta \tilde{t}_{ij} = \tilde{T}\left(\btt_i\right) - \tilde{T}\left(\btt_j\right) =  \frac{D_{\Delta t}}{c} \left[\ttau\left(\btt_i\right) - \ttau\left(\btt_j\right)\right] \eqqcolon \frac{D_{\Delta t}}{c}  \Delta \ttau_{ij} \ ,
	\label{tTD}
\end{equation}
where the SPT-modified Fermat potential is defined by
\begin{equation}
	\ttau\left(\btt\right) = \frac{1}{2} \left[ \btt - \bhb\left(\btt\right) \right]^2 - \tpsi\left(\btt\right) = \frac{1}{2} \left|\bta\left(\btt\right)\right|^2 - \tpsi\left(\btt\right) \ .
	\label{tFerma}
\end{equation}

\begin{figure}
	\resizebox{\hsize}{!}{\includegraphics{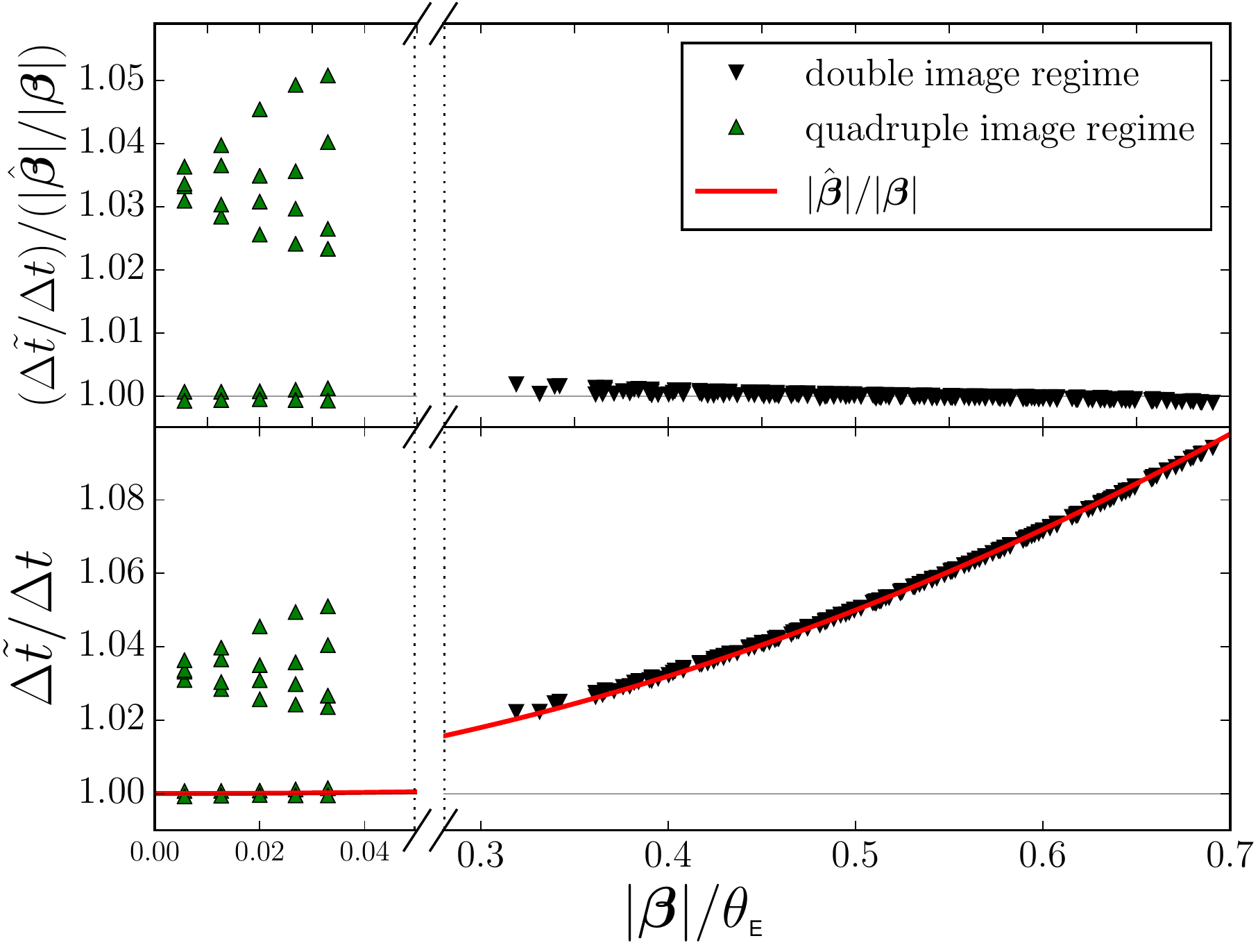}}
     	\caption{	Time delay ratios of image pairs between the NISg and the corresponding SPT-modified model. 
			The model parameters are $(\theta_{\text{c}}, \gamma_{\text{p}}) = (0.1\,\theta_{\text{E}},0.1)$ and the radial stretching is characterized
			by $f_2 = 0.4$.
			\emph{Top}:	$\Delta \tilde{t} / \Delta t$ normalized by the ratio $|\bhb|/|\bb|$ is close to $1$ in the double image regime and for opposite images in the
						quadruple image regime.
			\emph{Bottom}:	the impact of the SPT on the time delays is around a few percent, reaching a maximum of $12\%$ for the particular case of a source located almost on the 
						radial caustic curve but still inside $\mathit{B}_2$.
			} 
       	\label{figure:timedelays_NISg}       
\end{figure}

\begin{figure} 	
	\resizebox{\hsize}{!}{\includegraphics{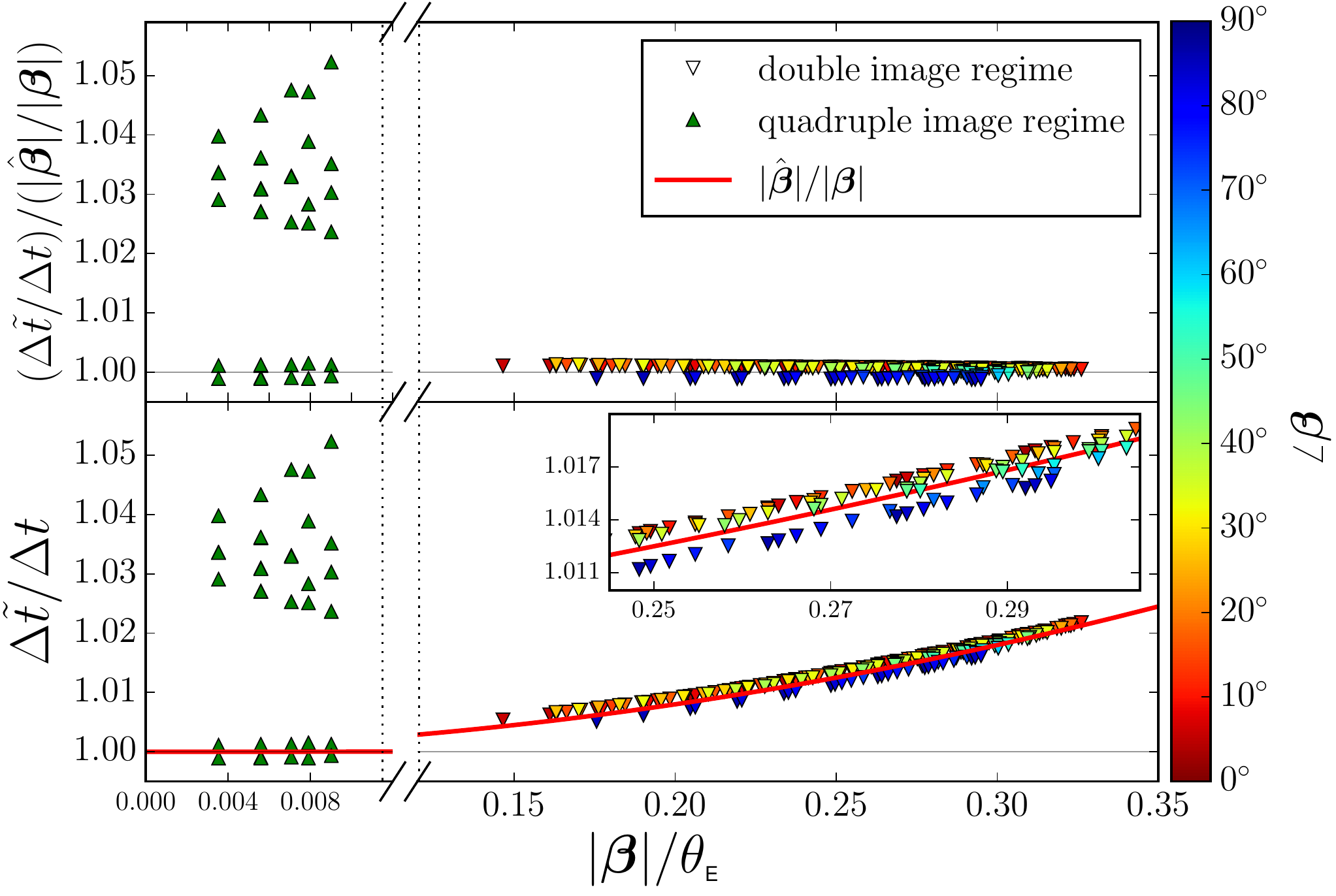}	}
     	\caption{	Time delay ratios of image pairs between the NIE and the corresponding SPT-modified model. 
			The model parameters are $(\theta_{\text{c}}, \epsilon) = (0.1\,\theta_{\text{E}},0.15)$ and the radial stretching is characterized
			by $f_2 = 0.4$. The time delay ratios in the quadruple (resp. double) image regime are depicted with triangles (resp. inverted triangles). 
			The red line shows the source position ratios $|\bhb|/|\bb|$ and the color-coding refers to the azimuth angle $\angle \bb$. 
			\emph{Top}:	even though the dispersion is slightly larger compared to the quadrupole model, $\Delta \tilde{t} / \Delta t$ normalized by the ratio $|\bhb|/|\bb|$ is still close to $1$ in 
						the double image regime and for opposite images in the quadruple image regime.
			\emph{Bottom}:	the time delay ratio dispersion is clearly related to $\angle \bb$ (see the text for more details). 
						The impact of the SPT on the time delays is also around a few percent, reaching a maximum of around $5\%$ for a particular combination of non-opposite images
						of a source located inside $\mathit{B}_1$. 
			}
       	\label{figure:timedelays_NIE}       
\end{figure}

We present here the representative results obtained for two classes of models: the quadrupole NISg as defined in Eqs. \eqref{quadrupole} 
and a non-singular isothermal elliptical lens (NIE). The NIE surface mass density $\kappa$ is intrinsically non-axisymmetric and is defined by \citep[see e.g.][]{Keeton_MassCatalog_2001}
\begin{equation}
	\kappa(\bt) = \frac{\tE}{2 \sqrt{\tc^2 + \rho^2}} \ ,
	\label{kappa_NIE}
\end{equation} 
where the variable $\rho$, constant on ellipses with axis ratio $q = \sqrt{(1-\epsilon)/(1+\epsilon)}$, is characterized by
\begin{equation}
	\rho = \sqrt{\frac{\theta_{x}^2}{1 - \epsilon} + \frac{\theta_{y}^2}{1 + \epsilon}} \ .
	\label{rho_NIE}
\end{equation} 
For the rest of this section, the quadrupole model parameters are set to $(\tc, \sub{\gamma}{p}) = (0.1\,\tE, 0.1)$ and the NIE model parameters to $(\tc, \epsilon) = (0.1\,\tE, 0.15)$. 
These models are deformed by an SPT corresponding to a radial stretching defined in Eq.\,\eqref{radialstretch} with $f_0 = 0$ and $f_2 = 0.4$.
In both cases, we used \pythonpackage\ to create mock images $\bt$ of two separated grids of sources $\bb$, which cover together the corresponding regions $\mathit{B}_1 \cup \mathit{B}_2$. The size and shape of 
$\mathit{B}_1 \cup \mathit{B}_2$ are defined by both the model and SPT parameters, hence differ from the quadrupole to the NIE (see Fig.\,\ref{figure:Delta_theta_map_sourceplane} for the quadrupole 
and Fig.\,\ref{figure:Delta_theta_map_sourceplane_NIE} for the NIE). Making use of Eq.\,\eqref{tilde_alpha_simplified}, we compute the corresponding images $\btt$ of the SPT-modifed 
sources $\bhb$. We finally derive the time delay $\Delta t_{ij}$ and $\Delta \tilde{t}_{ij}$ and represent their ratios in Figs. \ref{figure:timedelays_NISg} (for the NISg) and \ref{figure:timedelays_NIE} (for the NIE). 
Because of their similarities, we discuss the NISg and the NIE simultaneously.

In the double image regime, 
the time delay ratios scale almost like the ratios $|\bhb|/|\bb|$. 
The color-coding in Fig.\,\ref{figure:timedelays_NIE} refers to the azimuth angle of $\bb$, denoted as $\angle \bb$.
Even though the dispersion is slightly larger for the NIE, the deviations from $|\bhb|/|\bb|$ do not exceed $0.5\%$ in all cases.
The inset in Fig.\,\ref{figure:timedelays_NIE} clearly shows that the dispersion of the time delay ratios is the effect of the relative direction of $\bb$ with respect to the 
orientation of the axis of the elliptical iso-density contours (here equal to $0^{\circ}$). The deviations from $|\bhb|/|\bb|$ are maximum for $\angle \bb = 0^{\circ}$
and $\angle \bb = 90^{\circ}$, and minimum for $\angle \bb \approx 45^{\circ}$. 
A similar behavior is observed for the NISg, but with respect to the orientation of the external shear (also equal to $0^{\circ}$).
We suggest that Eq.\,\eqref{hTDratio1}, valid for the axisymmetric case (see Sect.\,\ref{subsection:axisym:hatTD}), may also be valid in the non-axisymmetric case for sufficiently large values of $|\bb|$,
\begin{equation}
	\frac{\Delta \tilde{t}}{\Delta t} \approx 1 + f(|\bb|) = \frac{\left|\bhb(\bb)\right|}{|\bb|} \ .
	\label{tTDratio_NISg}
\end{equation}
Actually, even for the most unfavorable cases, Eq.\,\eqref{tTDratio_NISg} provides at least a fairly good estimate of $\Delta \tilde{t} / \Delta t$.
It turns out that these two particular examples are representative of the time delay ratio behavior for double image configurations produced by a non-axisymmetric lens.
Thus, the impact of the SPT in the double image regime comes mainly from the ratios $|\bhb|/|\bb|$, in the same way as for the axisymmetric case. In particular, the largest time delay 
ratios $(\Delta \tilde{t}/\Delta t)_{\text{max}}$ (considering only the double image configurations for now) is obtained for the source position 
$\bb \in \mathit{B}_2$ characterized by the largest radial coordinate $|\bb|$ and denoted as $\sub{\bb}{max}$. 
Therefore, $(\Delta \tilde{t}/\Delta t)_{\text{max}}$ depends on $\sub{\bb}{max}$ and the latter depends on both the deformation function and the original lens model 
parameters, which define the size of $\mathit{B}_2$.
For the NISg model depicted in Fig.\,\ref{figure:timedelays_NISg}, we find $\subexp{\bb}{max}{NISg} \approx (0.562, 0.413)\,\tE$, $\left|\subexp{\bb}{max}{NISg}\right| \approx 0.697\,\tE$, leading to 
$(\Delta \tilde{t}/\Delta t)_{\text{max}}^{{\rm NISg}} \approx 1.121$, i.e., an impact of around $12\%$ on $H_0$. 
For the NIE model depicted in Fig.\,\ref{figure:timedelays_NIE}, we find $\subexp{\bb}{max}{NIE} \approx (0.335, 0.0)\,\tE$, $\left|\subexp{\bb}{max}{NIE}\right| \approx 0.335\,\tE$, leading to 
$(\Delta \tilde{t}/\Delta t)_{\text{max}}^{{\rm NIE}} \approx 1.028$, i.e., an impact of around $3\%$ on $H_0$. Similarly to the axisymmetric case, the impact of the SPT on time delays may substantially vary 
according to the nature of the original lens model.

A different behavior is observed for the case of the quadruple image regime. 
As first pointed out in SS13 from a pure empirical case, the time delay ratios of image pairs between the original and 
SPT-modified models are not conserved, i.e., $(\Delta \tilde{t} / \Delta t)_{ij} \neq (\Delta \tilde{t} / \Delta t)_{ik}$ with $i < k \leq 4$. For this reason, even though only $3$ independent time delays can be obtained from a 
quadruple image configurations, we represent in Figs.\,\ref{figure:timedelays_NISg} and \ref{figure:timedelays_NIE} the time delay ratios for all six image permutations $(i,j) \in [(1,2), (1,3), (1,4), (2,3), (2,4), (3,4)]$.
We note that the criterion chosen for ordering the images ($1$ to $4$) is the extra light travel time, from smallest to largest. The pair of opposite images, namely $(\bt_1,\,\bt_2)$ and $(\bt_3,\,\bt_4)$, leads 
to $\Delta \tilde{t} / \Delta t$ close to $1$, regardless of the adopted original lens model we have tested (see the pairs of green triangles close to $1$ in bottom panels in Fig.\,\ref{figure:timedelays_NISg} 
and \ref{figure:timedelays_NIE}). Owing to the symmetry of the image configurations $(\mathit{B}_1 \ni \bb \sim \boldsymbol{0})$, 
the time delays $\Delta t_{12}$ and $\Delta t_{34}$ are smaller than the time delays between other image combinations, tending towards $0$ when $\bb$ approaches $\boldsymbol{0}$. The same holds true for the 
SPT-modified time delays, while we note that $\Delta \tilde{t}_{12} \approx \Delta t_{12}$ and $\Delta \tilde{t}_{34} \approx \Delta t_{34}$. For sources $\bb \in \mathit{B}_1$, the mean impact of the 
SPT, denoted as $\langle \Delta \tilde{t} / \Delta t \rangle$, is around of a few percent for both the NISg and the NIE. In contrast to the NISg, the impact of the SPT for the case of the NIE is larger in the 
quadruple image regime than in the double image regime. 
This only reflects that $1 + f\left(\left|\subexp{\bb}{max}{NIE}\right|\right) \lesssim \langle \Delta \tilde{t} / \Delta t \subexp{\rangle}{}{\text{NIE}}$ while 
$1 + f\left(\left|\subexp{\bb}{max}{NISg}\right|\right) \gg \langle \Delta \tilde{t} / \Delta t \subexp{\rangle}{}{\text{NISg}}$.
 
\subsection{The alternative SPT-modified time delays}
\label{subsection:alternativetimedelays}

\begin{figure}
	\centering
   	\includegraphics[height=6.3cm]{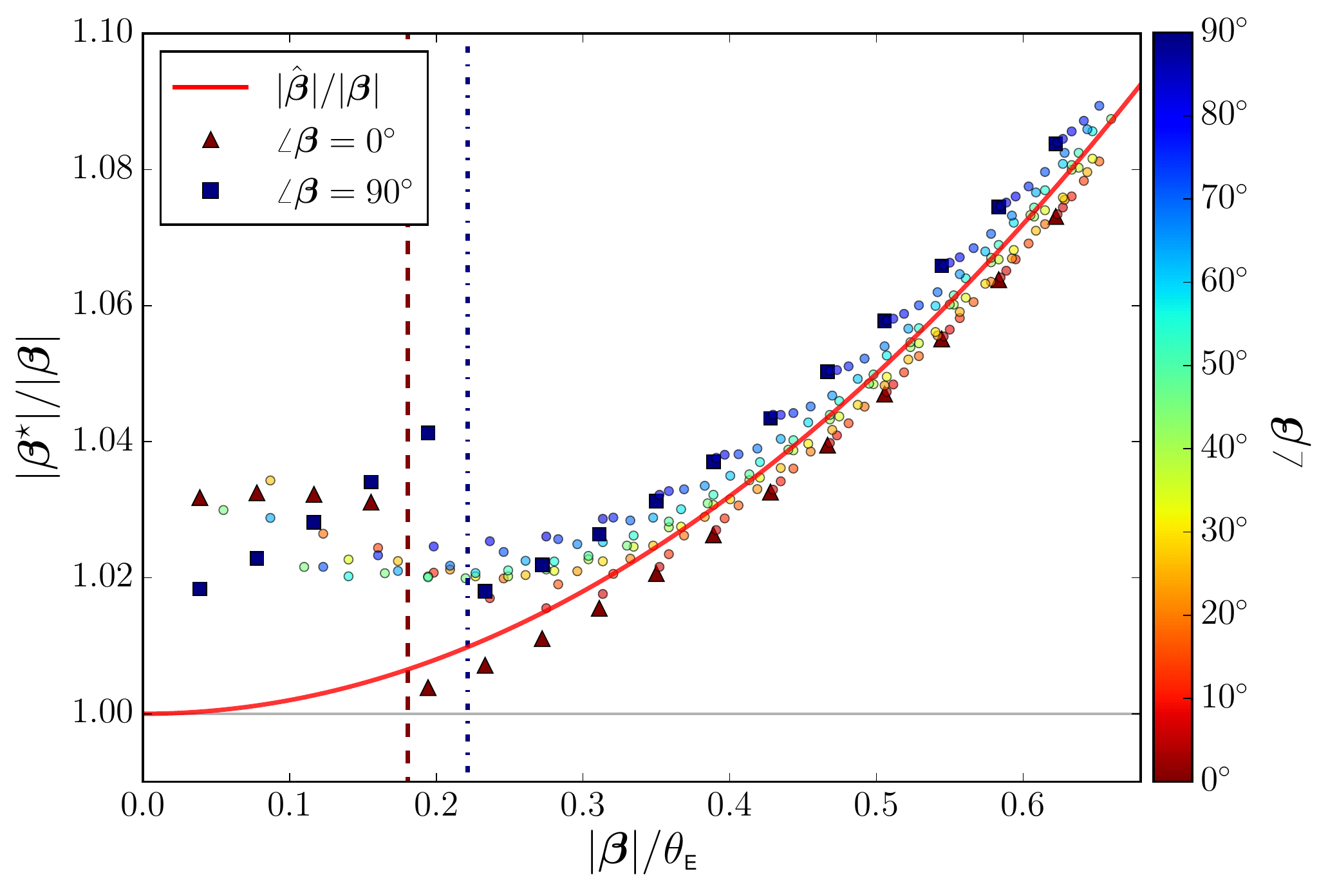}
     	\caption{	Ratios $|\bbs|/|\bb|$ plotted against $\bb$ in units of $\theta_{\text{E}}$. The source positions $\bbs$ are obtained by means of a numerical optimization of the cost function $h(\bbs)$ 
			defined in Eq.\,\eqref{costfunction}. The resulting source mapping $\bbs(\bb)$ is slightly anisotropic: $|\bbs|$ is larger than $|\bhb|$ for $\angle \bb > 45^{\circ}$ and smaller for 
			$\angle \bb < 45^{\circ}$. The color-coding refers to the azimuth angle $\angle \bb$ in the source plane. When a source crosses the tangential caustic curve, the mapping shows $\bbs(\bb)$ 
			discontinuities. Two particular jumps are highlighted: (1) sources depicted with triangles pass by the cusp located on the $\beta_{x}$-axis (the corresponding $\beta_{x}/\theta_{\text{E}}$ is 
			identified by the dashed vertical line); and (2) sources depicted with squares pass by the cusp located on the $\beta_{y}$-axis (the corresponding $\beta_{y}/\theta_{\text{E}}$ is 
			identified by the dash-dotted line.)
			}
       	\label{figure:tilde_beta_star_radial}       
\end{figure} 

\begin{figure}
	\centering
   	\includegraphics[height=14.3cm]{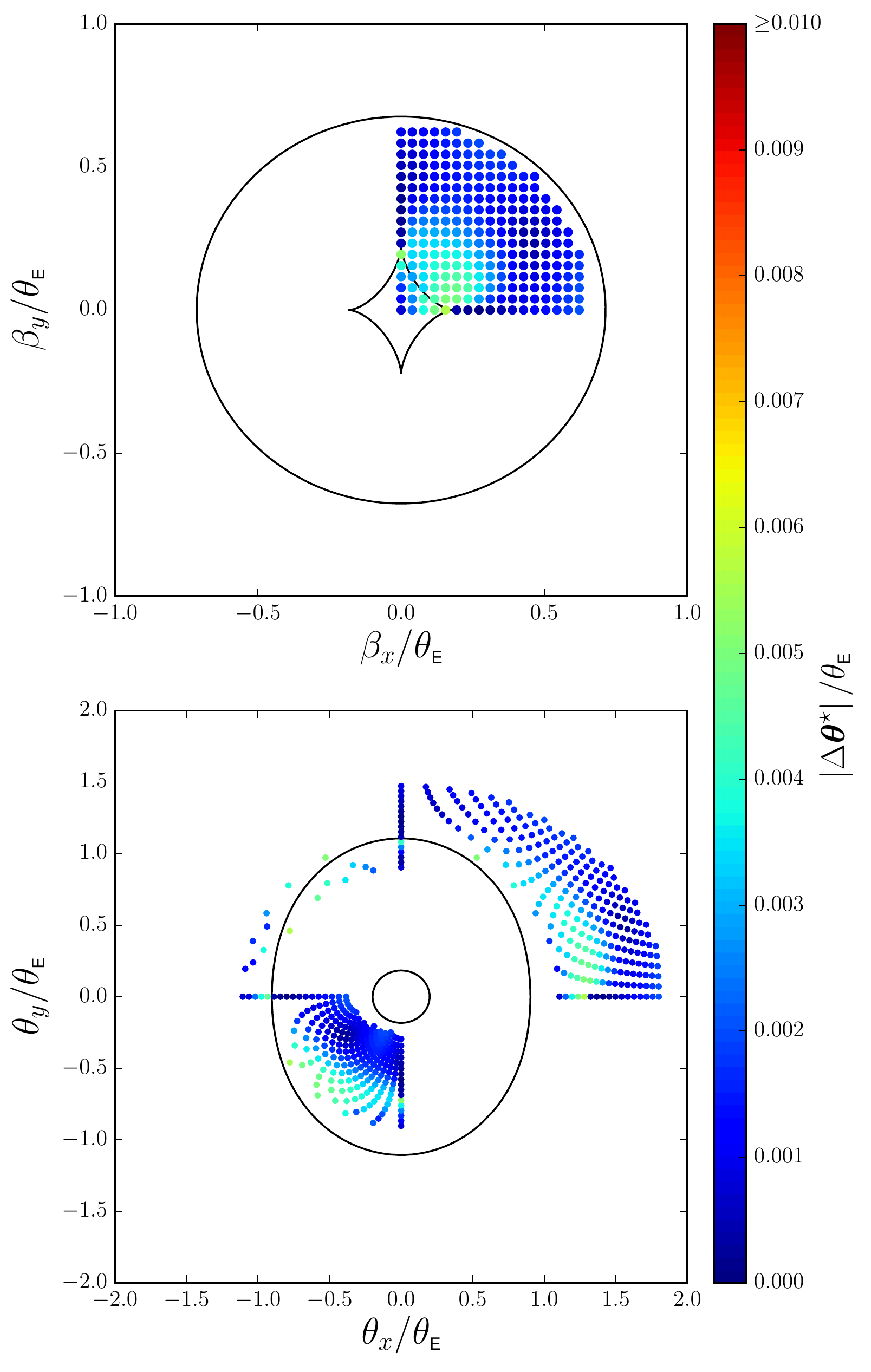}	
     	\caption{	\emph{Bottom}:	set of mock lensed images $\bt$ produced by the source positions shown in the \emph{top panel} and lensed by an NIS plus external shear characterized by $\tc = 0.1\,\theta_{\text{E}}$ and 
						$\gamma_{\text{p}} = 0.1$	. The color-coding represents the image offsets $|\Delta \bts|$ in units of $\theta_{\text{E}}$ between $\bt$ and $\bts$. The latter are the images of the source 
						positions $\bbs$ that are lensed by the SPT-modified lens associated with $\bta$. The sources $\bbs$ result from the numerical optimization of the cost function $h(\bbs)$. Most of the new image
						configurations are now characterized by $|\Delta \bts| \leq \varepsilon_{\text{acc}}$.
			\emph{Top}: 	grid of source positions $\bb$ covering the radial range $0 \leq |\bb| \leq 0.66\,\theta_{\text{E}}$ in the first quadrant of the source plane. Each source produces a set of lensed images $\bt$ 
						(shown in the \emph{bottom panel}) under the original deflection law $\ba$ and the corresponding set of $\bts$ under the curl-free deflection law $\bta$. The color-coding refers to the largest offset 
						$|\Delta \bts|_{\text{max}} \equiv h(\bbs)$ associated with each source $\bbs$. The solid black curves locate the caustic curves (\emph{top panel}) and the critical curves 
						(\emph{bottom panel}), respectively, for the NIS plus external shear.}
       	\label{figure:tilde_images_star}       
\end{figure}

In Sect.\,\ref{subsection:validity}, we have shown that the capability of the deflection law $\bta$ 
to predict the same multiple images as predicted by $\ba$ (with an accuracy of $\sub{\varepsilon}{acc}$) is very limited (see Eq.\,\ref{DeltaTheta}). Based upon a representative example, 
Fig.\,\ref{figure:Delta_theta_map_sourceplane} shows that only a very small region $(\mathit{B}_1)$ in the source plane leads to indistinguishable quadruple image configurations. 
In this section, we investigate a method to extend the region in the lens plane where the offsets $|\Delta \bt|$ are smaller than $\sub{\varepsilon}{acc}$. The idea consists in finding 
source positions $\bbs$ in the vicinity of $\bhb(\bb)$ that lead to new image positions $\bts = \bbs + \bta\left(\bts\right)$ in such a way that the offsets $|\Delta \bts| \coloneqq |\bts - \bt|$ 
are as small as possible. The search for each $\bbs$ is based on the numerical minimization of the cost function $h(\bbs)$ defined by
\begin{equation}
	h(\bbs) = \text{max}\left(\left|\bts_i\left(\bbs\right) - \bt_i\left(\bhb\right) \right|\right) \ ,
	\label{costfunction}
\end{equation} 
using the Levenberg-Marquardt algorithm \citep{algo_LM}. Because we expect $\bbs$ to be close to the corresponding $\bhb$, we always choose the latter as first guesses
while we do not restrict $\bbs$ to share the same direction as $\bhb$. Thus, the resulting source mapping $\bbs(\bhb)$, hence $\bbs(\bb)$, may not be isotropic as for the 
radial stretching.

As we shall see, this approach may drastically increase the region $\mathit{B}_1 \cup \mathit{B}_2$ (in particular $\mathit{B}_1$) while benefiting from a simple implementation. 
However, we must point out that this method suffers several flaws. First, the source mapping $\bbs(\bb)$ lacks a solid analytical basis. While $\bhb(\bb)$ gives rise to $\bha$ which is 
analytically connected to $\bta$ by definition\footnote{We also recall the analytical relation $\bn \cdot \bta = \bn \cdot \bha = 2\,\hkp\ $ over $\mathcal{U}$.}, there is no apparent link 
between $\bta$ and $\bbs$. Moreover, the way it is obtained precludes any further analytical investigation. Secondly, there is no definitive guarantee for the source mapping $\bbs(\bb)$ to 
be one-to-one over $\mathcal{U}$. Finally, a successful minimization of the cost function $h$ for a given $\bbs$ does not guarantee $|\Delta \bts|$ to be smaller than $\sub{\varepsilon}{acc}$.
Indeed, the solution $\bbs$ only corresponds to the one for which the cost function $h(\bbs)$ is the smallest in the vicinity of $\bhb$, being potentially larger than $\sub{\varepsilon}{acc}$. 
This is particularly true for sources which are located very close to the caustic curves.
At least, we have $|\Delta \bts| \leq |\Delta \bt|$ where the equality holds when $\bbs = \bhb$. For these reasons, we point out that this numerical approach cannot supplant the analytical 
reconsideration of how the curl-free deflection law $\bta$ is defined. However, the combination of $\bbs(\bb)$ and $\bta$ constitutes a physically meaningful alternative to $\bb$ and $\ba$, 
and deserves to be considered. 

To illustrate the method, we adopt the same lens model and SPT as in Sect\,\ref{subsection:validity}. We also consider the same grid of sources $\bb$ covering 
the first quadrant of the source plane and restricted to $0 \leq |\bb| \leq 0.66\,\tE$. We illustrate the results of the numerical optimization in Figs.\,\ref{figure:tilde_beta_star_radial} and 
\ref{figure:tilde_images_star}. The ratio $|\bbs|/|\bb|$ plotted against $|\bb|/\tE$ in Fig.\,\ref{figure:tilde_beta_star_radial} clearly shows the slight anisotropy of the source mapping $\bbs(\bb)$
resulting from the numerical optimization. Indeed, for original sources located on a quarter circle with a radius $|\bb|$, the corresponding $|\bbs|$ depend on the azimuth angle $\angle \bb$; $|\bbs|$ is 
larger than $|\bhb|$ for $\angle \bb > 45^{\circ}$ and smaller for $\angle \bb < 45^{\circ}$. 
For sources leading to double image configurations, the $|\bbs|$ scatter is around $1\%$. 
Furthermore, discontinuities in the mapping $\bbs(\bb)$ appear when a source crosses the tangential caustic curve. Two particular jumps are highlighted for sources passing through the two cusps. The one 
located on the $\beta_{x}$-axis (resp. $\beta_{y}$-axis) is depicted in Fig.\,\ref{figure:tilde_beta_star_radial} by the dashed line (resp. dash-dotted line).
Due to these discontinuities, an extended source which crosses the tangential caustic curve is not mapped smoothly onto an SPT-modified extended source. This effect propagates to the image plane, but the impact 
on the corresponding lensed image is not observable, as shown in Fig.\,\ref{figure:tilde_images_star}.
Similarly to Fig.\,\ref{figure:tilde_images}, the bottom panel in Fig.\,\ref{figure:tilde_images_star} represents the image positions $\bt$ and the color-coding shows the offsets $|\Delta \bts|$ in unit of $\tE$.
The same color-coding is also applied to the sources where only the largest offset between the corresponding pairs of lensed images are shown (see top panel in Fig.\,\ref{figure:tilde_images_star}). 
Figure\,\ref{figure:tilde_images_star} shows that almost all the offsets $|\Delta \bts|$ are now smaller than $\sub{\varepsilon}{acc}$, even for sources located inside the tangential caustic curve. 
It is worth stating that a finer source grid would have led to a larger number of sources located very close to the caustic curves, for which the optimized cost function may be larger than $\sub{\varepsilon}{acc}$. 
Compared to Fig.\,\ref{figure:tilde_images}, some image positions depicted in Fig.\,\ref{figure:tilde_images_star} show an offset $|\Delta \bts|$ (after the optimization process) larger than the offset $|\Delta \bt|$ 
(before the optimization process). For example, the image position $\bt_k = (1.242,0)\,\tE$ of the source $\bb = (0.117,0)\,\tE$ is characterized by $|\Delta \bt_k| = 0.001\,\tE$ while $|\Delta \bts_k| = 0.005\,\tE$.
This behavior stems from the fact that, for a given $n$-image configuration, the optimization process minimizes only the largest offset but not all the $n$ offsets simultaneously (because of the ${\rm max}(\cdot)$ function in 
Eq.\,\ref{costfunction}). Thus, while the largest offset becomes smaller, the offset $|\Delta \bts_k|$ also varies during the optimization process, leading at the end to $|\Delta \bt_k| < |\Delta \bts_k| \leq h(\bbs)$.

\begin{figure}
	\centering
	\includegraphics[height=7.9cm]{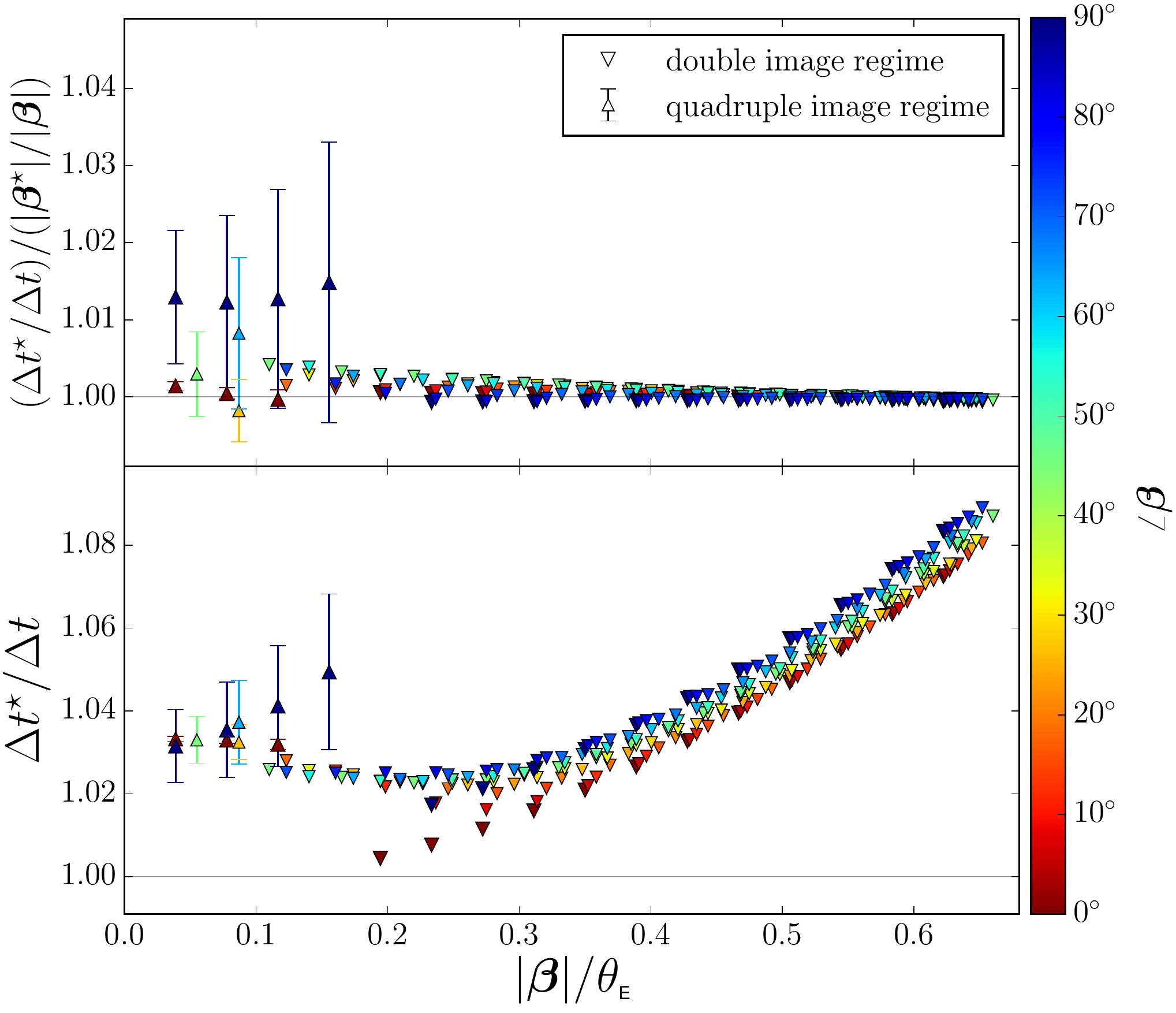}	
     	\caption{	Time delay ratios of image pairs between the NISg and the corresponding SPT-modified model associated with the curl-free deflection field $\bta$. 
			The model parameters are $(\theta_{\text{c}}, \gamma_{\text{p}}) = (0.1\,\theta_{\text{E}},0.1)$ and the radial stretching defining the source positions $\bhb$ is characterized by $f_2 = 0.4$.
			The time delays $\Delta t_{ij}^{\star}$ are evaluated for pair of images $\bts_i(\bbs)$ and $\bts_j(\bbs)$. Properties of the source positions $\bbs$ are shown in Fig.\,\ref{figure:tilde_beta_star_radial}.
			\emph{Top:}	$\Delta t^{\star} / \Delta t$ normalized by the ratio $|\bbs|/|\bb|$ is close to $1$ in the double image regime and for all combination $i$ and $j$ of images $\bt$, in the
						quadruple image regime, when the azimuth angle $\angle \bb = 0^{\circ}$. As $\angle \bb$ increases, the time delay ratios $\Delta t^{\star} / \Delta t$ deviates from $|\bbs|/|\bb|$ 
						and the corresponding $\sigma(\Delta t^{\star} / \Delta t)$ also increases.
			\emph{Bottom:}	the impact of the SPT on the time delays is around a few percent, reaching around $8\%$ for source positions $\bb$ located almost on the radial caustic curve.}
       	\label{figure:timedelays_general}       
\end{figure}

Now that we have obtained a large set of indistinguishable image configurations, we derive the time delays $\Delta t_{ij}$ and $\Delta t_{ij}^{\star}$ between image pairs where $\Delta t_{ij}^{\star}$ is defined by
\begin{equation}
	\Delta t_{ij}^{\star} = \tT\left(\bts_i\right) - \tT\left(\bts_j\right) = \frac{D_{\Delta t}}{c} \left[\ttau\left(\bts_i\right) - \ttau\left(\bts_j\right)\right] \ . 
	\label{TDstar}
\end{equation}
Figure\,\ref{figure:timedelays_general} plots the time delay ratios $\Delta t^{\star} / \Delta t$ (bottom panel) and the time delay ratios normalized by $|\bbs|/|\bb|$ (top panel), both against $|\bb|/\tE$. 
The same color-coding as in Fig.\,\ref{figure:tilde_beta_star_radial}  is also applied to the time delay ratios. 
In the double image regime, 
the normalized time delay ratios $(\Delta t^{\star} / \Delta t)/(|\bbs|/|\bb|)$ is close to $1$. This suggests once again that the time delay ratios scale almost like the source 
ratios $|\bbs|/|\bb|$, even though the source mapping is not perfectly isotropic. Thus, the equation 
\begin{equation}
	\frac{\Delta t^{\star}}{\Delta t} \approx \frac{\left|\bbs(\bb)\right|}{|\bb|} 
	\label{starTDratio}
\end{equation}
still holds for the slight anisotropic source mapping $\bbs(\bb)$ and is particularly true for large $|\bb|$ inside the radial caustic curve.
In the quadruple image regime, the time delay ratios of image pairs between the models are also not perfectly conserved. In either panels in Fig.\,\ref{figure:timedelays_general}, we represent the mean between the six 
time delay ratios and the corresponding standard deviation, denoted as $\sigma(\Delta t^{\star} / \Delta t)$ and depicted with error bars. However, $\sigma(\Delta t^{\star} / \Delta t)$ 
is around a minimum of $0.001$ for $\angle \bb = 0^{\circ}$ when $\bb$ points towards the same direction as the external shear. Conversely, it reaches a maximum of $0.02$ for $\angle \bb = 90^{\circ}$ 
when $\bb$ points perpendicularly to the direction of the external shear. In addition for $\angle \bb = 0^{\circ}$, the time delay ratios $\Delta t^{\star} / \Delta t$ scale like 
$|\bbs|/|\bb|$, regardless of the image combinations. Thus, as the azimuth angle $\angle \bb$ increases, the time delay ratios $\Delta t^{\star} / \Delta t$ deviates from $|\bbs|/|\bb|$ and the corresponding 
$\sigma(\Delta t^{\star} / \Delta t)$ also increases. We recall that prior to any numerical optimization,
the time delays ratios $\Delta \tilde{t} / \Delta t$ 
were close to $1$ only for pairs of opposite images. Quantitatively, the impact of the SPT on the time delays agrees with what we have obtained in Figs.\,\ref{figure:timedelays_NISg} and \ref{figure:timedelays_NIE}, 
that is an impact of a few percent. It turns out that most of the conclusions drawn in Sect.\,\ref{subsection:timedelaysforvalid} seem to apply in this case as well.

In Sect.\,\ref{subsection:validity} we suggest to reconsider the criterion for the validity of an SPT as formulated in \citet[][]{SPT_USS17}. This shall lead to
more appropriate definition of the SPT-transformed deflection law and potential, enlarging the scope where the SPT is a valid invariance transformation. 
Because the Eq.\,\eqref{starTDratio} holds within the numerical approach described in this section, we expect that Eq.\,\eqref{tTDratio_NISg} will also remain true 
when considering new definitions for $\bta$ and $\tpsi$, and an isotropic SPT. Thus, most of the conclusions drawn in this paper should remain unchanged.

\section{Conclusions}
\label{section:conclusions}

The SPT corresponds to a global invariance transformation of the source plane. It gives rise to a new deflection law, which leaves almost all the lensing observables invariant. We have 
studied the impact of the SPT on the time delays between pairs of lensed images produced by both axisymmetric and non-axisymmetric lenses. Although we have mainly considered the case of 
an isotropic SPT described by a radial stretching of the form $\bhb = [1 + f(|\bb|)]\,\bb$, we have also addressed a particular case for which the STP is slightly anisotropic.

Owing to its simplicity, we were able to deal with the axisymmetric case in an analytical way. We have shown that the time delay ratios of image pairs between the SPT-modified and original models
approximate as the ratios between the SPT-modified and original source positions, namely $\Delta \hat{t} / \Delta t \approx \hb / \beta$. Based on simple analytical arguments, we have demonstrated
that this relation holds for any axisymmetric lenses and even deformation function $f$. In particular, this relation is exact when the SPT reduces to an MST or when the lens is described by 
an SIS model, regardless of the deformation function $f$. For an NIS model deformed by a radial stretching of the form $\hb(\beta) = [1 + f_2\,\beta^2 / (2\,\tE^2)]\,\beta$, 
we have shown that $\Delta \hat{t} / \Delta t$ deviates no more than a few hundredth of percent from $\hb / \beta$. In addition, we have demonstrated that the source mapping can be expressed in 
terms of the mean surface mass densities, that is $\hb/\beta = (1 - \langle \hkp \rangle)/(1 - \langle \kappa \rangle)$. 

Quantitatively, the impact of the SPT on time delays may reach a few percent for particular image configurations, and depends on various factors.
Indeed, $\hb$ depends on the choice made for the deformation function $f$ and its corresponding deformation parameters. 
Not all combinations of an original mass profile $\kappa$, a function $f$, and a set of deformation parameters yield a physically meaningful SPT-modified mass profile $\hkp$. It means that
the parameter validity range of a given SPT, hence the validity range of $\hb$ or $\langle \hkp \rangle$, needs to be studied on a case-by-case basis. 

When we drop the axisymmetry assumption for the original lens model, the SPT-modified deflection angle $\bha$ is not a curl-free field caused by a mass distribution corresponding to a gravitational lens (SS14).
An alternative deflection field $\bta$ was proposed in USS17, namely the closest curl-free approximation to $\bha$ which fulfills the validity criterion $|\Delta \ba(\bt)| \equiv |\bta(\bt) - \bha(\bt)| \leq \sub{\varepsilon}{acc}$ over a region 
$\mathcal{U}$ in the lens plane where multiple images occur. We have studied in detail the relevance of this criterion using $\sub{\varepsilon}{acc} = 5 \times 10^{-3}\,\tE$ as suggested in USS17. We have shown that this criterion is not appropriate, 
in particular for positions close to the critical lines. Indeed, we have demonstrated to first order that the offsets $|\Delta \bt|$ between the original and SPT-modified lensed images depends on the SPT-modified Jacobi matrix whose
impact become larger as we get closer to the critical curves, $|\Delta \bt| \approx \left|\hat{\mathcal{A}}^{-1}(\bt)\,\Delta \ba(\bt) \right|$. Thus, the criterion $|\Delta \ba(\bt)| \leq \sub{\varepsilon}{acc}$ over $\mathcal{U}$ does not guarantee 
the image offsets $|\Delta \bt|$ to be smaller than $\sub{\varepsilon}{acc}$ over the same region. In those cases, the deflection field $\bta$ produces image configurations which can be observationally distinguished from the
original ones. As a result, we suggest that the criterion $|\Delta \ba(\bt)| \leq \sub{\varepsilon}{acc}$ proposed in USS17 should be reconsidered, which also means a revision of how the curl-free deflection law $\bta$ and $\tpsi$ are defined. 

For the indistinguishable image configurations produced by $\bta$, we have studied how the time delays are affected by the SPT. For a quadrupole (NIS + shear) and a NIE models, we have shown that, once again, the time delay ratios
scale like the source position ratios, i.e. $\Delta \tilde{t} / \Delta t \approx |\bhb| / |\bb|$, when two images are produced. This result holds for opposite image pairs when four images are produced. For other image pair combinations, we 
confirm that the time delay ratios are not conserved. Thus, accurate time delay ratios measurements should help to reduce the degeneracy between SPT-generated models.
However, the impact of the SPT remains low with deviations not larger than a few percent for the illustrative examples we have considered. 

To extend the range of indistinguishable image configurations produced by $\bta$, we have slightly modified the source mapping by means of a numerical optimization, $\bhb(\bb) \rightarrow \bbs(\bb)$. 
We tested this method on a quadrupole model (NIS + shear) deformed by a radial stretching. The new source mapping differs from a radial stretching in the sense that $\bbs(\bb)$ is slightly anisotropic and 
shows discontinuities when a source crosses the tangential caustic curve. In the same way as for the previous cases, the time delay ratios of image pairs scale like the source position 
ratios, i.e. $\Delta t^{\star} / \Delta t \approx |\bbs| / |\bb|$,
when two images are produced. When four images are produced, the time delay ratios are sensitive to the azimuth angle of $\bb$. We have shown that they scale like the source position ratios when the azimuth angle 
is parallel to the external shear direction and deviate to a few percent when the azimuth angle is perpendicular to the external shear direction. 
As a general conclusion, the impact of the SPT on time-delay cosmography seems not be as crucial as initially suspected, leading to deviations that do not exceed a few percent.

In a future work, we aim to reconsider the validity criterion in more detail by redefining the curl-free deflection field $\bta$ and the corresponding potential $\tpsi$.
Although new definitions for $\bta$ and $\tpsi$ will affect the time delays, we expect that it will not modify substantially the results presented in this paper, likely leaving 
most of the conclusions unchanged.

\begin{acknowledgements}
We thank Dominique Sluse and Sandra Unruh for useful discussions. 
This work was supported by the Humboldt Research Fellowship for Postdoctoral Researchers. 

\end{acknowledgements}

\bibliographystyle{aa}
\bibliography{bibtex_GL}

\begin{appendix} 
\section{Proof that $\varepsilon_{\text{AB}} = 0$ for an SIS and a radial stretching of the form \eqref{radial_stretching_1D}}
\label{appendix:epsilonAB_SIS}

For the sake of clarity, we first recall the definition of $\sub{\varepsilon}{AB}$ given in Eq.\,\eqref{EpsilonAB} 
\begin{equation}
	\sub{\varepsilon}{AB} =  \int_{|\tB|}^{\tA} \beta(\theta) \left[f(\beta(\theta)) - f(\varbeta)\right]\ \text{d}\theta \ .
	\label{EpsilonABbis}
\end{equation}
In this section, we will proof that $\sub{\varepsilon}{AB} = 0$ for an SIS and for any deformation function $f(\beta)$ which satisfies a few reasonable conditions.  
By definition, $f(\beta)$ must be even to preserve the symmetry and $1 + f(\beta) + \beta\,\text{d}f(\beta)/\text{d}\beta > 0$ guarantees the mapping to be one-to-one.
Let us assume that $f(\beta)$ is a real analytic function, its Maclaurin series expansion 
thus exists and is simply given by
\begin{equation}
	f(\beta) = \sum_{n=0}^{+\infty} f_{n}\ \frac{\beta^{n}}{n!} \ ,
	\label{maclaurin_f}	
\end{equation} 
where $f_n \coloneqq \text{d}^{n}f/\text{d}\beta^n$ is evaluated in $\beta = 0$ and $f_{2k+1} = 0$ for all $k \in \mathbb{Z}^{*}$ to preserve the symmetry. 
Because of the linearity of integration and recalling the lens equation $\beta(\theta) = \theta - \tE \theta / |\theta|$ for an SIS, Eq.\,\eqref{EpsilonABbis} transforms into
\begin{equation}
	\sub{\varepsilon}{AB} =  \sum_{n=0}^{+\infty} \frac{f_{n}}{n!} \left( \int_{|\tB|}^{\tA} (\theta - \tE)^{n+1}\ \text{d}\theta \right) - f(\varbeta) \int_{|\tB|}^{\tA} (\theta - \tE)\ \text{d}\theta  \ ,
	\label{EpsilonABbis_maclaurin}
\end{equation}
where $n$ is now a positive even integer or $0$.
Keeping in mind that $n+2$ is even, the linearity of the integrand guarantees that
\begin{equation}
	\int_{|\tB|}^{\tA} (\theta - \tE)^{n+1}\ \text{d}\theta = \left[ \frac{(\theta - \tE)^{n+2}}{n+2} \right]_{|\tB|}^{\tA} = 0 \ ,
	\label{EpsilonABbis_I1}
\end{equation}
where in the last step we used $\tA - \tE = \varbeta$ and $|\tB| - \tE = - \varbeta$. 
Because of the latter expression is valid for all positive even integer $n$ and for $n = 0$, we deduce that all the successive terms in the series vanish, as the second integral in Eq.\,\eqref{EpsilonABbis_maclaurin}, 
leading to $\sub{\varepsilon}{AB} = 0$. 
Furthermore, the accuracy of \eqref{maclaurin_f} is not affected by the value at which the function is evaluated as long as the series converges. Thus, we finally require that the radius 
of convergence $r$ of the Maclaurin series \eqref{maclaurin_f} satisfies the condition $r \geq \tE$. If satisfied, this very plausible assumption assures the series expansion \eqref{maclaurin_f} to be 
exact for all $\beta < \tE$. 

\section{Proof that $\varepsilon_{\text{AB}} \approx 0$ for any axisymmetric model and radial stretching of the form \eqref{radial_stretching_1D}}
\label{appendix:epsilonAB_general}

\begin{figure}
	\centering
   	\includegraphics[height=7.35cm]{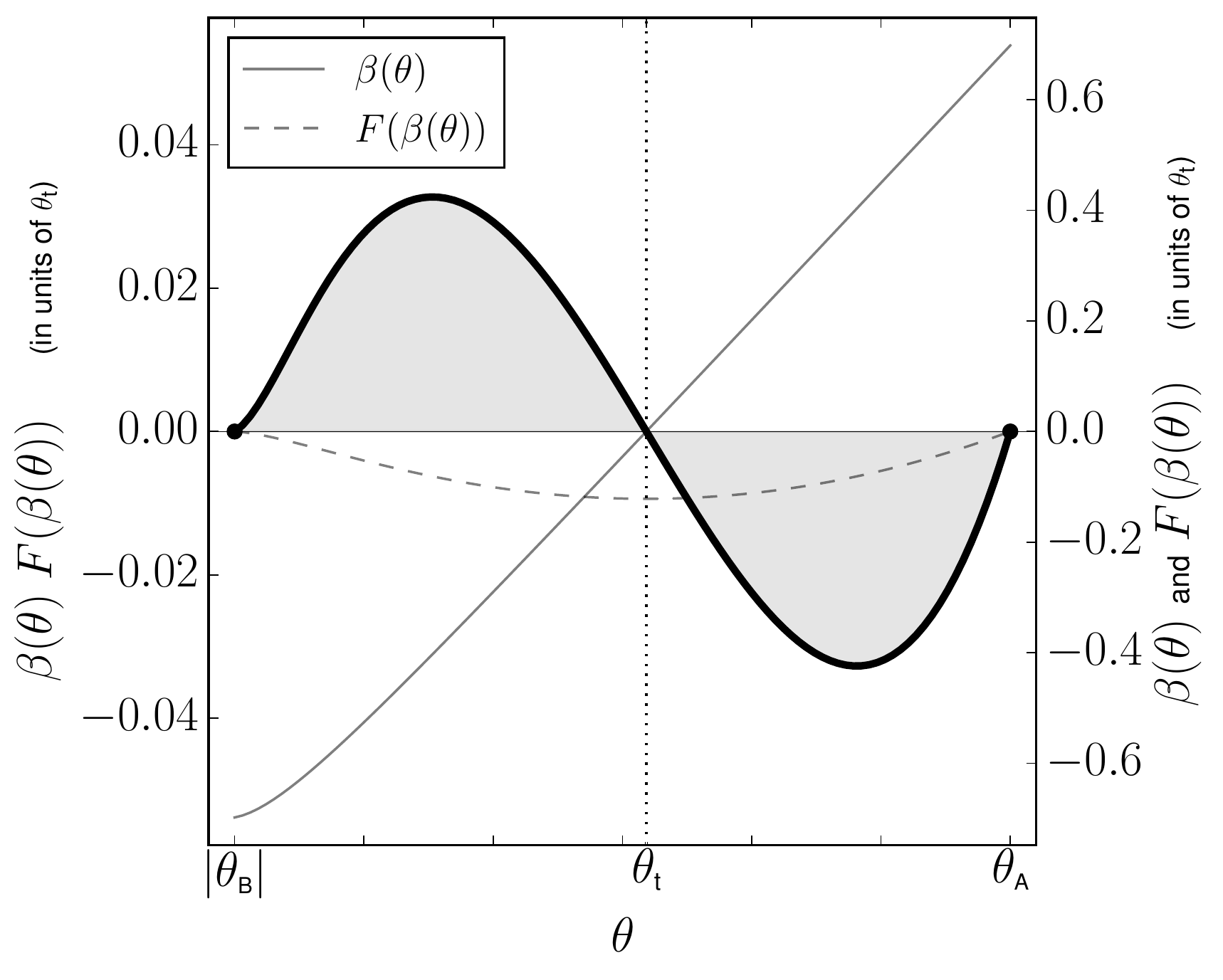}
     	\caption{	Graph of the integrand $\beta(\theta) F(\beta(\theta))$ (thick black curve) defined in Eq.\,\eqref{EpsilonAB} for the `unfavorable' case $|\theta_{\text{B}}| \rightarrow \theta_{\text{r}}$. This integrand is composed of the functions 
			$\beta(\theta)$ (solid gray curve) and $F(\beta(\theta))$ (dashed gray curve). The (almost) symmetrical graph of $\beta(\theta)$ with respect to $(\ttt, 0)$ and $F(\beta(\theta))$ with 
			respect to the axis $\theta = \ttt$ (dotted vertical line) implies the (almost) symmetry of the integrand with respect to $(\ttt, 0)$. As a consequence, the two gray areas (almost) compensate 
			each other implying that $|\varepsilon_{\text{AB}}| \ll \ttt$, hence $|\eta_{\text{AB}}| \ll 1$, and confirms the validity of the Eq.\,\eqref{hTDratio1}.}
       	\label{figure:integrand}       
\end{figure} 

To convince the reader, we consider separately the cases when $\varbeta$ tends to $0$ ($\varbeta \rightarrow 0$) and when $\varbeta < \beta(\tr)$, where $\tr$ corresponds to the angular 
radius of the radial critical curve. 

For $\varbeta \rightarrow 0$, we see from Eq.\,\eqref{EpsilonAB} that the integration interval degenerates into $\{ \ttt \}$, where $\ttt$ corresponds to the angular 
radius of the tangential critical curve. Thus, the integrand also tends to 0 and $\eAB \rightarrow 0$. In addition, $\eAB \rightarrow 0$ faster than $\Delta \tau_{\text{\scalebox{.9}{AB}}} \rightarrow 0$ due 
to the first term in Eq.\,\eqref{oTD2} while $1 + f(\varbeta) \rightarrow 1$. Thus, we deduce that the $\eAB$-term in Eq.\,\eqref{hTDfinal} can be neglected when $\varbeta \rightarrow 0$.

For $\varbeta < \beta(\tr)$, we define the quantity 
\begin{equation}
	\nAB = \frac{\sub{\varepsilon}{AB}/\Delta \sub{\tau}{AB}}{\hb(\varbeta)/\varbeta}  \ ,
	\label{EtaAB}
\end{equation}
and show that $|\nAB| \ll 1$, which is sufficient to guarantee that 
the $\eAB$-term in Eq.\,\eqref{hTDfinal} can be neglected. With this aim in mind, we analyze the graph of the integrand $\beta(\theta) \left[f(\beta(\theta)) - f(\varbeta)\right] \eqqcolon \beta(\theta) F(\beta(\theta))$ defined 
in Eq.\,\eqref{EpsilonAB} and for which an example is shown in Fig.\,\ref{figure:integrand}. 
Based on general considerations, we expect the graph of the integrand to be almost symmetric with respect to the point $(\theta_{\text{t}}, 0)$ over the interval $[|\tB|, \tA]$. 
First, the graph of $\beta(\theta)$ 
monotonically increases over $[|\tB|, \tA]$ and always crosses the $\theta$-axis at the position $\theta = \theta_{\text{t}} \approx (\tA + |\tB|)/2$. 
These statements stem from the general properties of axisymmetric lenses in the case of a single lens plane \citep[see e.g. the section 3.1 in][]{Schneider_Saas-Fee}. In addition, because of $\mbox{d}\beta(\theta)/\mbox{d}\theta$ is 
almost constant $(\approx 1)$ for most $\theta \in [|\tB|, \tA]$, the graph of $\beta(\theta)$ is almost symmetric with respect to the point $(\theta_{\text{t}}, 0)$. The largest asymmetry occurs for 
$\varbeta \rightarrow \beta(\theta_{\text{r}})$ for which $|\tB| \rightarrow \theta_{\text{r}}$ (this `unfavorable' case is actually the one shown in Fig.\,\ref{figure:integrand}). 
Secondly, the graph of $F(\beta(\theta))$ 
can only cross the $\theta$-axis at the positions $\tA$ and $\tB$ over $[|\tB|, \tA]$, and is almost symmetric 
with respect to the axis $\theta = \ttt$. 
We easily confirm that $\tA$ and $\tB$ are $\theta$-intercepts from $F(\beta(\tA)) = F(\beta(\tB)) = F(\varbeta) = 0$. To show there exists 
no other $\theta-$intercept within $[|\tB|, \tA]$, we use a reductio ad absurdum argument. If the graph of $F(\beta(\theta))$ crosses the $\theta$-axis at a third position $\tI \in [|\tB|, \tA]$, then $F(\beta(\tI)) = 0$ implies that 
$\tI$ corresponds to the position of a lensed image of the source $\varbeta$. Since the third image $\tC$ always satisfies the condition $0 < |\tC| < |\tB| < \tI < \tA$, hence $\tC \notin [|\tB|, \tA]$, the existence of this fourth lensed 
image violates the so-called `odd number theorem' \citep{Dyer_Roeder_1980, Burke_1981} in the case of a single lens plane. 
In addition, because of the symmetry of $\beta(\theta)$ for $\theta \in [|\tB|, \tA]$ and recalling that $f$ is an even function of $\beta$, we have $F(\beta(\ttt + \delta \theta)) \approx F(\beta(\ttt - \delta \theta))$, with 
$\delta \theta \in [0, \text{min}(\tA - \ttt, \ttt - |\tB|)]$, which guarantees $F(\beta(\theta))$ to be almost symmetric with respect to the axis $\theta = \ttt$.
Combining all these statements leads to the conclusion that the integrand $\beta(\theta) F(\beta(\theta))$ is also almost symmetric with respect to $(\ttt, 0)$. Thus, the integral \eqref{EpsilonAB} consists in differencing the 
(almost identical) gray areas displayed in Fig.\,\ref{figure:integrand}, which tends to compensate each other, leading to $|\eAB| \ll \ttt$. Finally, from Eq.\,\eqref{oTD2} we deduce that $|\Delta \tau_{\text{\scalebox{.9}{AB}}}| \approx 2\,\varbeta\,\ttt\ $ 
due to the symmetry of $\beta(\theta)$ over $[|\tB|, \tA]$. As a conclusion, we find $|\nAB| \ll 1$ from Eq.\,\eqref{EtaAB}, which confirms that the $\eAB$-term in Eq.\,\eqref{hTDfinal} can be in general neglected.

\end{appendix}

\end{document}